\documentclass[prd,twocolumn,nofootinbib,aps,tightenlines,preprintnumbers,notitlepage,longbibliography,superscriptaddress]{revtex4-1}

\usepackage[dvipsnames]{xcolor}
\usepackage{color}
\usepackage{tikz}
\usepackage[caption=false]{subfig}
\usepackage{hyperref}
\usepackage{multirow}
\usepackage{siunitx}
\usepackage[export]{adjustbox}
\usepackage{graphicx}
\usepackage{dcolumn}
\usepackage{bm}
\usepackage[normalem]{ulem}
\usepackage{epstopdf}
\usepackage{tabularx}
\usepackage{suffix}
\usepackage{mathtools}
\newcommand{\bn}{\hat{\boldsymbol{n}}}

\usepackage{graphicx}
\usepackage{dcolumn}
\usepackage{bm}
\usepackage[normalem]{ulem}
\usepackage{xcolor}

\graphicspath{ {figures/} }

\usepackage{epstopdf}
\newcommand{\be}{\begin{eqnarray}}

\newcommand{\ee}{\end{eqnarray}}

\newcommand{\bbe}{\boldsymbol{e}}
\newcommand{\bv}{\boldsymbol{v}}

\newcommand{\br}{\mathbf{r}}
\newcommand{\bk}{\mathbf{k}}
\newcommand{\khat}{\hat{\mathbf{k}}}

\def\thetahat{\hat{\boldsymbol{\theta}}}
\def\phihat{\hat{\boldsymbol{\phi}}}
\def\bk{\boldsymbol{k}}

\def\spinup{\partial\kern-0.3em\raise0.42ex\hbox{\tiny\textbackslash}}
\def\spindown{\overline{\partial\kern-0.3em\raise0.42ex\hbox{\tiny\textbackslash}}}

\newcommand{\apjl}{Astrophys. J. Lett.}

\newcommand{\apjs}{Astrophys. J. Suppl. Ser.}

\newcommand{\mnras}{Mon. Not. R. Astron. Soc.}
\newcommand{\jcap}{J. Cosm. Astropart. Phys.}

\newcommand{\dd}{{\rm d}}

\newcommand{\bsk}{\bk}

\newcommand{\jhu}{William H. Miller III Department of Physics and Astronomy, 3400 N.\ Charles St., Baltimore, MD 21218, USA}

\newcommand{\nyu}{Center for Cosmology and Particle Physics, Department of Physics, New York University, New York, NY 10003}

\begin{document}

\title{Probing cosmic birefringence with polarized Sunyaev-Zel'dovich tomography}

\author{Nanoom~Lee}
\email{nanoom.lee@nyu.edu}
\affiliation{\nyu}

\author{Selim~C.~Hotinli}
\email{shotinl1@jhu.edu}
\affiliation{\jhu}

\author{Marc~Kamionkowski}
\email{kamion@jhu.edu}
\affiliation{\jhu}

\date{\today}

%%%%%%%%%%%%%%%%%%%%%%
%%%% Abstract%%%%%%%%%
%%%%%%%%%%%%%%%%%%%%%%

\begin{abstract}
    
    If the physics behind dark energy and/or dark matter violates the parity symmetry assumed in the standard cosmological paradigm, the linear polarization of the cosmic microwave background (CMB) photons can rotate due to their coupling to the dark sector. Recent $3\sigma$ hints of this ``cosmic birefringence" in the $EB$ spectrum of the CMB polarization motivates us to pursue new directions to independently validate and characterize the signal. Here, we explore the prospects to probe cosmic birefringence from small-scale fluctuations in the CMB using polarized Sunyaev-Zel'dovich (pSZ) tomography. We find that pSZ can be used to infer the redshift dependence of cosmic birefringence and also help calibrate the instrumental polarization orientation. To illustrate the prospects,  we show that pSZ tomography may probe an axion-like dark energy model with masses $m_\phi \lesssim 10^{-32}$eV with $\mathcal{O}(0.1)$ degrees of rotation between reionization and recombination.
    
\end{abstract}

\maketitle

%%%%%%%%%%%%%%%%%%%%%%
%%%% Intro %%%%%%%%%%%
%%%%%%%%%%%%%%%%%%%%%%

\section{Introduction}

The upcoming decade will see a great expansion in our ability to infer the fundamental characteristics of the Universe thanks to the next generation of CMB surveys, such as the ongoing Simons Observatory~\citep{Ade:2018sbj,SimonsObservatory:2019qwx}, the upcoming CMB-S4~\citep{Abazajian:2016yjj,Abazajian:2022nyh}, and  the futuristic CMB-HD~\citep{Sehgal:2019ewc,CMB-HD:2022bsz}; and galaxy surveys such as DESI~\citep{Aghamousa:2016zmz}, the Vera Rubin Observatory (VRO)~\citep{2009arXiv0912.0201L}, and the proposed MegaMapper \citep{Schlegel:2019eqc}. These surveys open a promising new window of opportunity for inference through using the CMB as a backlight and utilizing the small-scale non-Gaussian information in the CMB maps in cross-correlation with galaxy surveys. This program will provide the best sensitivity to a wide range of deviations from the standard cosmological paradigm~(see e.g.~Refs.~\citep{Dvorkin:2008tf,Dvorkin:2009ah, Smith:2016lnt, Kamionkowski:2008fp, Yadav:2009eb, Gluscevic:2009mm, Alizadeh:2012vy, Deutsch:2017cja, Deutsch:2017ybc, Meyers:2017rtf, Hotinli:2018yyc, Hotinli:2020ntd, Hotinli:2021hih, Hotinli:2020csk, Cayuso:2021ljq,Kumar:2022yus} for examples). Here, we focus on the polarized Sunyaev-Zel'dovich (pSZ) effect induced by the scattering of CMB photons off energetic electrons in the large-scale structure~\citep{1999MNRAS.310..765S,Sazonov:1999zp,Hu:2001kj} and evaluate the prospects to probe parity-violation due to an axionlike dark energy (or dark matter) which could change the remote-quadrupole signal that can be inferred from tomographic measurements of pSZ~\citep{2012PhRvD..85l3540A,Deutsch:2018umo,Deutsch:2017ybc,Deutsch:2017cja}.

A pseudoscalar field $\phi$ such as the axion can be a candidate for either dark energy or dark matter~\citep{Dine:1982ah,Abbott:1982af,Preskill:1982cy,Marsh:2015xka,Ferreira:2020fam} and rotate
the plane of linear polarization of photons as they travel through space via a Chern-Simons term in the Lagrangian density, $\mathcal{L}\in-(1/4)g\phi F_{\mu\nu}\tilde{F}^{\mu\nu}$~\citep{PhysRevD.37.2743}. Here, $F_{\mu\nu}$ is the electromagnetic tensor, and $g$ is the axion-photon coupling constant. This rotation (often referred to as ``cosmic birefringence'') would lead to a nonzero odd-parity $EB$ and $TB$ power-spectra of the CMB photons from recombination or reionization~\cite{Lue:1998mq} (see also recent review \cite{Komatsu:2022nvu}), as well as the secondary CMB polarization anisotropies, such as the pSZ signal, sourced at lower redshifts $z\lesssim6$, depending on when it occurs. Recently, there have been $\sim 3\sigma$ hints of this cosmic birefringence \cite{Minami:2020odp,Diego-Palazuelos:2022dsq,Eskilt:2022cff,Eskilt:2022wav} using Planck \cite{Planck:2018bsf,Planck:2020olo,Planck:2018lkk} and WMAP \cite{WMAP:2012fli} polarization data, by which this work is motivated.

\begin{figure}[t!]
\includegraphics[width=1\columnwidth, trim= 0 20 0 0]{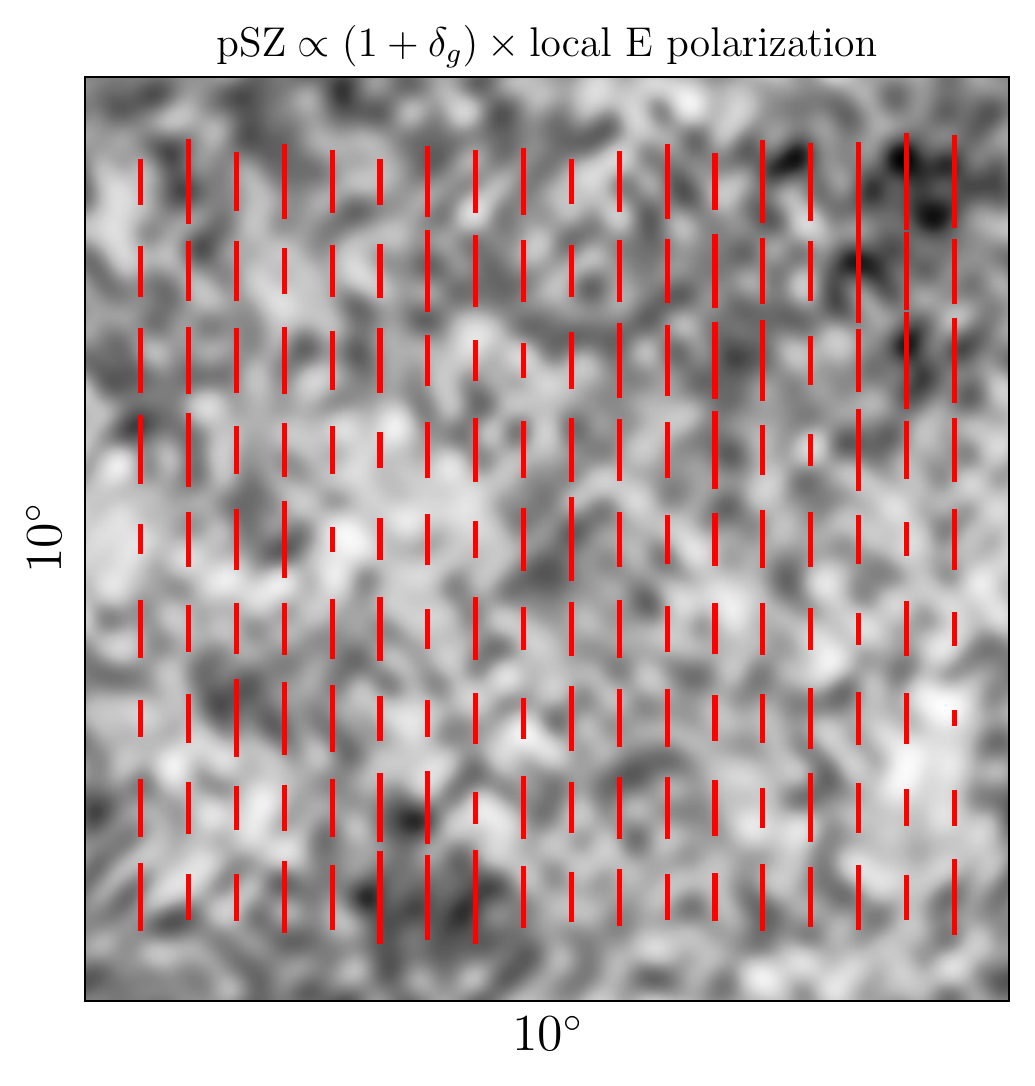}
\caption{Small-scale fluctuations of CMB polarization magnitude induced by galaxy (or electron-density) fluctuations are shown. If the background was homogeneous, one would see a uniform magnitude. For illustrative purposes, an artificial (not physical) background field is constructed and the variation in polarization magnitude is amplified for a clear visualization.}
\label{fig:illustration}
\end{figure}

Here we show that pSZ tomography~\citep{Deutsch:2018umo,Deutsch:2017ybc} allows measurement of the cosmic birefringence angle as a function of redshift on the light cone. The way this works is as follows: There is a CMB intensity quadrupole at any point in the Universe induced by the Sachs-Wolfe effect, just as there is a CMB quadrupole locally.  This quadrupole has long-range coherence and can be inferred with some fidelity from measurement of the CMB temperature/polarization fluctuations on the very largest scales (as may be done, e.g., by an experiment like LiteBird that maps the full sky).  Scattering of CMB photons by ionized electrons after reionization induces a CMB polarization whose {\it orientation} has long-range coherence but with small-scale  {\it magnitude} fluctuations induced by electron-density fluctuations on small scales.  These electron-density fluctuations correlate with the galaxy distribution.  Thus, by matching the small-scale polarization-magnitude fluctuations with a galaxy-redshift survey, the contribution to the polarization in any given direction can be inferred as a function of redshift (see Fig.~\ref{fig:illustration}).  Cosmic birefringence will lead to a rotation of the small-scale polarization fluctuations observed relative to the orientation expected from the largest-scale polarization pattern. The redshift information allows the degeneracy between the unknown calibration angle of CMB detectors~\citep{2009PhRvD..79j3002M,QUaD:2008ado} and the total rotation due to birefringence to be broken, as well as potentially distinguishing between models with different predictions for when rotation occurs in redshift. Here, we find pSZ tomography may improve the measurement of birefringence for future CMB experiments in a number of different ways.

This paper is organised as follows. We start in Sec.~\ref{sec:birefringence} by discussing the phenomenology of cosmic birefringence and describe the observable signature on the CMB maps. We briefly introduce the pSZ effect and the pSZ tomography in Sec.~\ref{sec:pSZ_tomo}, and demonstrate the effect of birefringence on the reconstructed remote quadrupole field. We introduce the experimental specifications we consider in our forecasts in Sec.~\ref{sec:forecast}. We present our results in Sec.~\ref{sec:results}. We conclude with a discussion in Sec.~\ref{sec:discussion}.

\section{Cosmic birefringence}\label{sec:birefringence}

\subsection{Phenomenology}

The cause of cosmic birefringence by a pseudoscalar field $\phi$ can be dark energy~\citep{Carroll:1998zi,Panda:2010uq} or dark matter~\citep{Finelli:2008jv,Fedderke:2019ajk}, depending on the effective mass of the $\phi$ field, $m_\phi^2\equiv\dd^2 V(\phi)/\dd\phi^2$. For masses lower than the Hubble constant today $m_\phi\lesssim H_0\simeq10^{-33}$eV, $\phi$ evolves slowly on its potential $V(\phi)$ and constitutes to a dark-energy-like component. Conversely if $m_\phi>H_0$, $\phi$ would have entered a regime of speedily evolving on its potential, and, for sufficiently large masses $m_\phi\gtrsim10^{-32}$eV, $\phi$ could approach, overshoot, and begin oscillating about a local minima of its potential, possibly constituting to a fraction of the dark matter today. The $\phi$ field is found to be viable candidate for \textit{all of dark energy} if $m_\phi\lesssim 10^{-33}$eV~\citep{Fujita:2020aqt} with $\Omega_\phi h^2=\Omega_\Lambda h^2\simeq0.69$. For the the mass range $10^{-33}{\rm eV}\leq m_\phi\leq 10^{-25}{\rm eV}$, the energy density of $\phi$ is constrained to be $\Omega_\phi h^{2}\leq0.006$ from the current large-scale structure and CMB measurements~\citep{Hlozek:2014lca} and if $m_\phi>10^{-28}$eV, $\phi$ would behave like dark matter~\citep{Hlozek:2014lca,Fujita:2020aqt}.

\begin{figure}[t!]
\includegraphics[width=1\columnwidth, trim= 0 30 0 0]{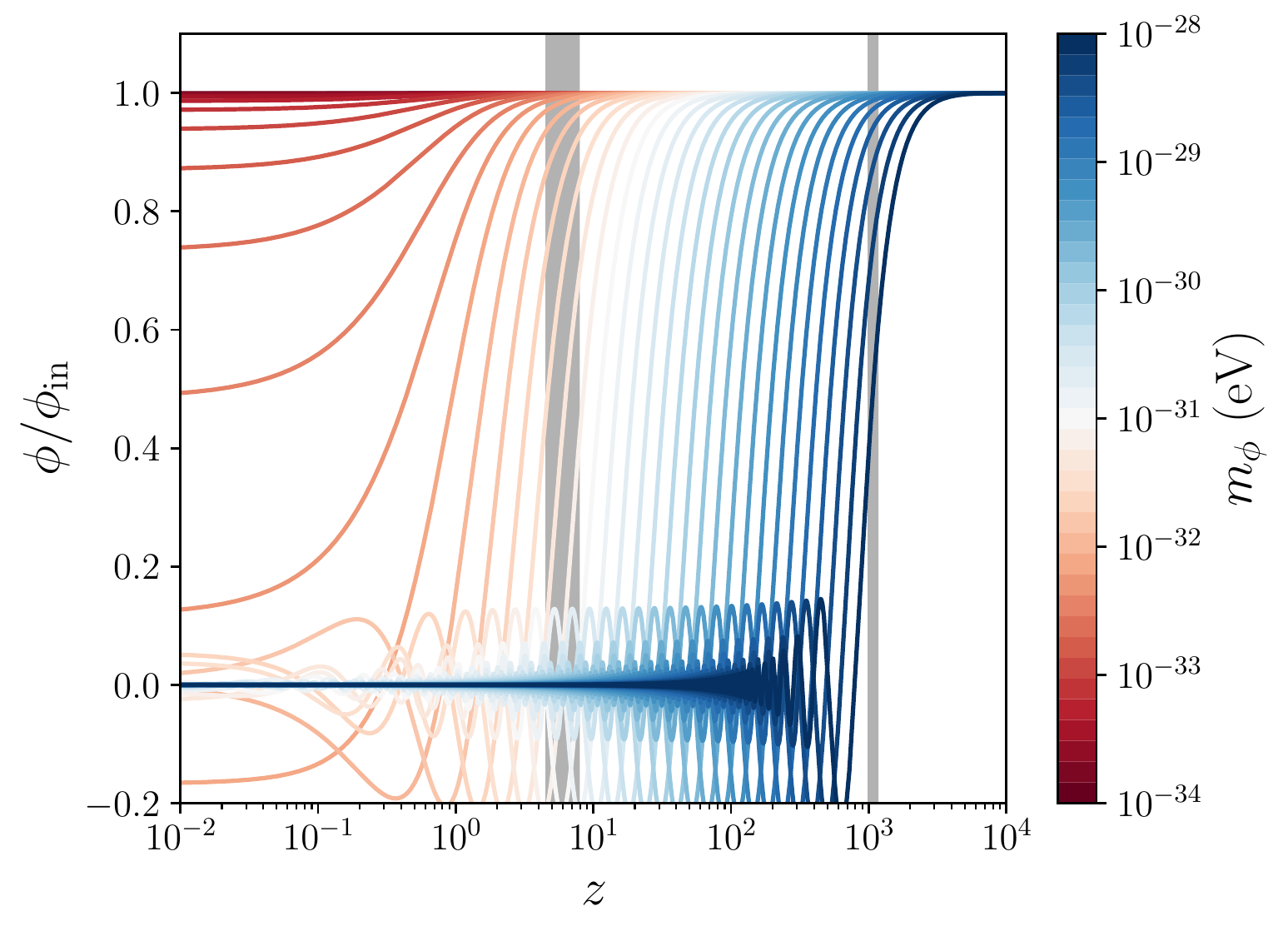}
\caption{The evolution of the $\phi$ field as a function of redshift $z$ for different masses $m_\phi$. The shaded regions correspond to epochs of reionization and recombination where the visibility function $g$ satisfy $g/\text{max}(g)>0.5$. The recombination visibility function is obtained using \textsc{hyrec}-2~\citep{Ali-Haimoud:2010hou,Lee:2020obi}, and for reionization we adopt the tanh model from Ref.~\citep{Lewis:2008wr}. Cosmic birefringence occurs when the $\phi$ field evolves significantly over its potential. For axion models that are dark-energy-like, the bulk of the rotation occurs at low redshifts after reionization.}
\label{fig:phi_evolution}
\end{figure}

We demonstrate the evolution of the $\phi$ field as a function of redshift and mass in Fig.~\ref{fig:phi_evolution}. Here, we have taken the equation of motion for $\phi$ as 
\be
\phi''+2\frac{a'}{a}\phi'+a^2m_\phi^2\phi=0\,,
\ee
where we set $V(\phi)=m_\phi^2\phi^2/2$ and the prime denotes the derivative with respect to the conformal time. Note that, however, our method can be applied to any kinds of potential in general including recent works studying axionlike dark energy \cite{Gasparotto:2022uqo,Choi:2021aze,Fujita:2020ecn,Nakagawa:2021nme}. Similar to Ref.~\citep{Nakatsuka:2022epj}, we take the $\phi$ field to be decoupled from the rest of the components in the Friedmann equation.\footnote{This approximation is valid only if the $\phi$ energy density is small or the mass is sufficiently small that $\phi$ behaves like dark energy.} Note that when $m_\phi\ll H(z)=a'/a^2$, the $\phi$ field does not evolve significantly on its potential. The birefringence $\alpha$ angle due to $\phi$ satisfies $\alpha=1/2g\int \dd t\, \dd\phi/\dd t$ where $\dd \phi/\dd t$ is the total derivative of $\phi$ along the trajectory of the photon. In what follows, we model $\alpha$ as sourced both from an axionlike $\phi$ field and also in a more general way to build intuition.

\subsection{Polarization parity-violation}

We model the effect of birefringence on the $E$- and $B$- mode polarization as 
\be
(Q'\pm iU')(\bn)=(Q\pm i U)(\bn)e^{\pm 2i\alpha}\,.
\ee
The modulation of the polarization modes due to birefringence caused between their emission and their arrival to our detectors can be written as 
\be
E_{\ell m}'&&=E_{\ell m}\cos2\alpha-B_{\ell m}\sin2\alpha \\ 
B_{\ell m}'&&=B_{\ell m}\cos2\alpha+E_{\ell m}\sin2\alpha\,.
\ee
The observed CMB spectra subject to cosmic birefringence are shown in Appendix~\ref{sec:app_1} with Eqs.~(A11)-(A15). 

\subsection{CMB calibration angle}

The measurement of the rotation angle $\alpha$ is understood to be subject to a systematic bias due to the miscalibration of polarization angles of polarization-sensitive detectors and other instrumental effects~\citep{2009PhRvD..79j3002M,QUaD:2008ado,2011ApJS..192...18K,2013ApJ...762L..23K}. 

This artificial parity-violation results in a nonzero $EB$ and $TB$ correlation equally on all multipoles and can potentially be \textit{surmounted} by taking advantage of the different $\ell$-dependence of the Galactic-foreground polarization emission where the latter is expected to be subject to negligible amount of birefringence~\citep{Minami:2019ruj}, for example. 

Alternatively, the systematic bias could be \textit{avoided} via a tomographic approach, comparing the birefringence experienced by the reionization and the recombination signals in the CMB~\citep{Sherwin:2021vgb,Nakatsuka:2022epj}.\footnote{It is also worthwhile to note that the consistent modelling of birefringence (as we do here) could lead to significant correlation between the birefringence experienced by recombination and reionization signals (due to correlated $\alpha$) which may result in less optimistic constraints compared to when the cross correlation is ignored.} The latter technique, however, is only sensitive to birefringence sourced between the reionization and recombination, while the method of using the Galactic-foreground is excepted to constrain the calibration angle at a precision $\sigma(\alpha_{\rm cal})\sim0.5^{\circ}$. 

In what follows we will remain model-agnostic about the effect of miscalibration when assessing the prospects to constrain the birefringence angle, considering a wide range of calibration-angle priors.

\section{Polarized Sunyaev-Zel'dovich tomography}\label{sec:pSZ_tomo}

\subsection{The polarized SZ effect}

The Stokes parameters $Q$ and $U$ that are measured in observations of CMB can be mapped onto a 2-sphere with unit vectors $\{\thetahat,-\phihat\}$ in the plane perpendicular to the line-of-sight direction $\bn$. The polarization vectors satisfy $\bbe_\pm(\bn)=(\thetahat\mp i\phihat)/\sqrt{2}$. This resulting complex CMB polarization along the line of sight takes the form,  
\be
(Q\pm iU)(\bn)=\frac{\sqrt{6}}{10}\int_{0}^{\chi_\star}\dd\chi \ \dot{\tau}(\chi\bn)e^{-\tau(\chi)}\,_{\pm}p(\chi\bn)\, , 
\ee
where $\chi$ is the comoving distance along our past light cone, $\chi = 0$ is today and $\chi_*$ is the comoving distance to the surface of last scattering. Here, $\tau$ is the optical depth of photons along the line of sight. The optical depth at redshift $z$ along $\bn$ can be written as $\tau(\chi\bn)=\int_0^\chi \dd\chi'\; \dd\tau/\dd\chi'(\chi'\bn)$ where $\dd\tau/\dd\chi(\chi\bn)=\sigma_Tan_e(\chi\bn)$ and $\sigma_T$ is the Thomson
cross section, $a$ is the scale factor, $n_e$ is the comoving free electron number density. The remote quadrupole field projected along the line of sight $\,_{\pm}p(\chi\bn)$ is given by
\be
\,_{\pm}p(\chi\bn) \equiv \sum_{m=-2}^{m=2} \Theta_{2m}(\chi\bn) \,_{\mp2}Y_{\ell m}(\bn)\label{eq:remote_quad}
\ee
where $\Theta_{2m}(\chi\bn)$ are the $\ell =2$ moments of the CMB temperature anisotropies observed at the position $\chi\bn$ on our past light cone.

The local CMB temperature quadrupole gets three contributions. These are the Sachs-Wolfe effect (SW), due to gravitational redshifting at the surface of last scattering; the integrated Sachs-Wolfe effect (ISW), resulting from the evolution of gravitational
potential; and the Doppler effect from the relative motion of electrons. The CMB quadrupole observed by an electron at the line-of-sight direction $\chi_e\bn_e$ satisfy
\be
\Theta_{2m}(\chi_e\bn_e)=\int\dd^2 \bn\; \Theta(\chi_e,\bn_e,\bn)Y^*_{2m}(\bn)\,,
\ee
where the contributions to the CMB temperature are
\be
\Theta(\chi_e,\bn_e,\bn)=&&\,\,\Theta_{\rm SW}(\chi_e,\bn_e,\bn)\nonumber\\
&&+\Theta_{\rm ISW}(\chi_e,\bn_e,\bn)\nonumber\\
&&+\Theta_{\rm Doppler}(\chi_e,\bn_e,\bn)\, \label{eq:Theta}.
\ee
Following Refs.~\citep{Deutsch:2017cja,Deutsch:2017ybc} we rewrite the local CMB temperature quadrupole as
\be
\Theta_{2m}(\chi_e\bn_e) &=& \int \frac{\dd^3k}{(2\pi)^3} \Psi_i(\bm{k}) T(k) [ \mathcal{G}_{\rm SW}+\mathcal{G}_{\rm ISW} + \mathcal{G}_{\rm Doppler}]\nonumber\\ 
&&\times Y_{2m}^*(\khat)\;e^{i\chi_e\bsk \cdot \bn_e},\label{eq:Theta_2m}
\ee
where $\Psi_i(\bm{k})$ is the primordial fluctuations and $\mathcal{G}$'s are the kernels of contributions to CMB temperature quadrupole. Then, the quadrupole field transfer function $\Delta^p_\ell(\chi_e, k)$ is given by
\be
\Delta_\ell^p(\chi_e,k) &=& -5i^\ell \sqrt{\frac{3}{8}} \sqrt{\frac{(\ell+2)!}{(\ell-2)!}} \frac{j_\ell(k\chi_e)}{(k\chi_e)^2} T(k) \nonumber \\
&& \times [ \mathcal{G}_{\rm SW}+\mathcal{G}_{\rm ISW} + \mathcal{G}_{\rm Doppler}],
\ee
and the angular power spectrum is given by
\be
\Big\langle p^{E*}_{\ell m}(\chi_e)\;p^E_{\ell'm'}(\chi_e')\Big\rangle = C_\ell^{p^E}(\chi_e,\chi_e')\delta_{\ell\ell'}\delta_{mm'},
\ee
where
\be
p^E_{\ell m}(\chi)= \int \frac{\dd^3k}{(2\pi)^3} \Delta_\ell^p(\chi, k) \Psi_i(\bsk) Y_{\ell m}^*(\khat).
\ee
For the detailed definitions and expressions see Appendix \ref{sec:app_c} and Ref.~\citep{Deutsch:2017cja,Deutsch:2017ybc} for derivations.

\begin{figure}[t!]
\centering
\includegraphics[width=\linewidth]{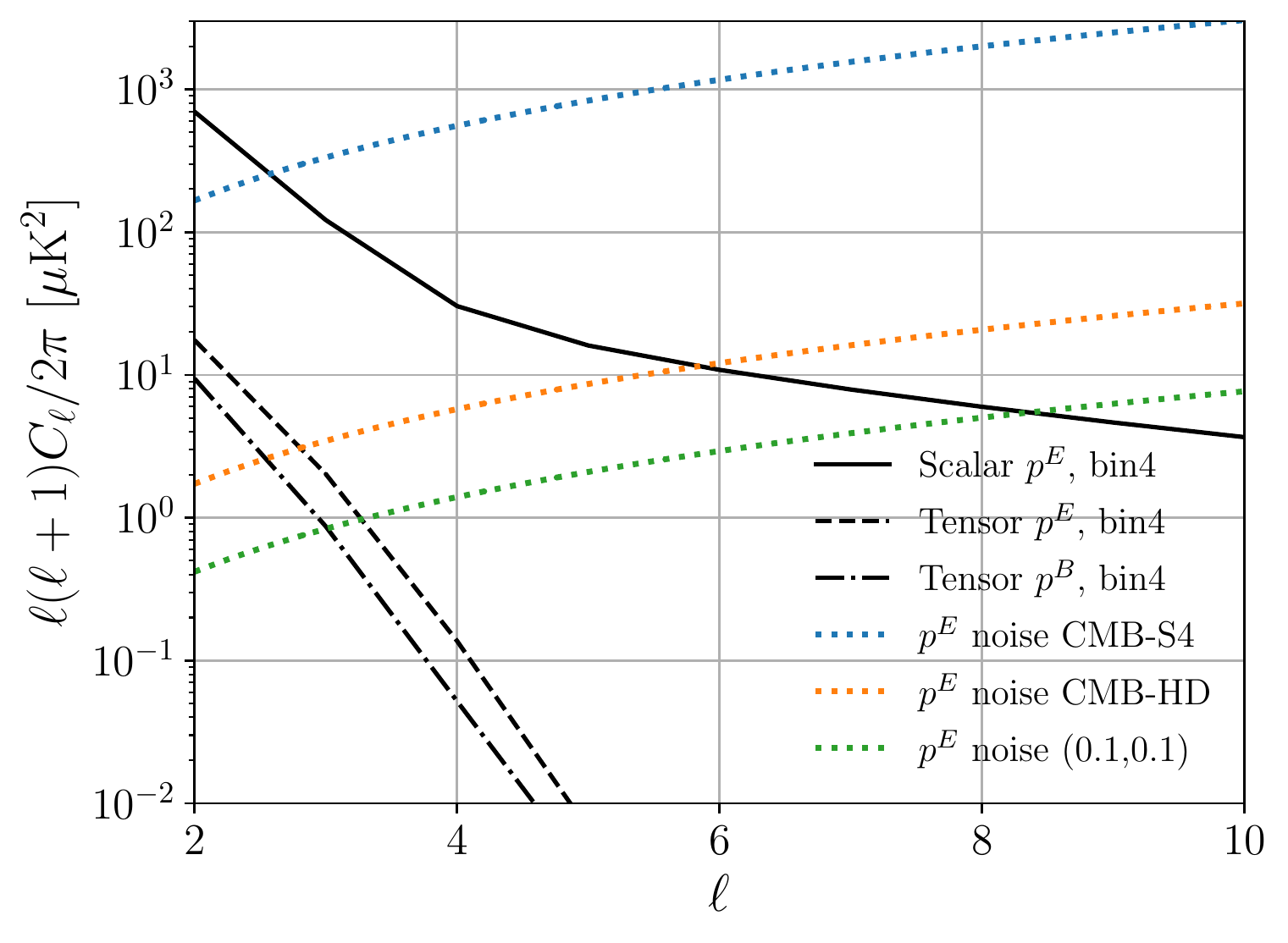}
\caption{The scalar and tensor remote $E$- and $B$- parity quadrupole signals. The tensory contributions (dashed and dash-dotted lines) correspond to tensor-to-scalar ratio of $r=0.05$. The dotted colored lines correspond to the anticipated reconstruction noises for pSZ tomography [using a CMB-S4- and a CMB-HD-like survey, as well as a futuristic survey, latter satisfying $(\Delta_T,\theta_{\rm FWHM})=(0.1 \mu{\rm K}',0.1')$] from our fourth redshift bin of which the median redshift is $z=0.76$, as described in the text. The pSZ tomography can utilize the auto- and cross-correlations of the reconstructed $E$- and $B$- mode remote quadrupole fields as well as the CMB temperature and polarization on the largest scales to probe parity-violating $EB$ and $TB$ spectra as a function of redshift.}
\label{fig:pedagogy}
\end{figure}

While we provide and use the expression for the remote CMB quadrupole originated only by the scalar primordial perturbations, it gets contributions from the tensor primordial perturbations as well, if any (see \citep{Deutsch:2018umo} for tensor contributions). We demonstrate the $E$- and $B$- mode quadrupole spectra in Fig.~\ref{fig:pedagogy} with tensor and scalar contributions. Throughout the rest of this work, we will assume no primordial tensor fluctuations.

\subsection{Effect of birefringence on the remote quadrupole}

The remote quadrupole field satisfy
\be
{\,_\pm}\hat{p}(\bn,\eta)\sim\left[(Q(\bn)\pm iU(\bn))^{\mathrm{obs}}\delta\tau(\bn,\eta)^{\mathrm{obs}}\right]_f\,,
\ee
where we defined fluctuations of the optical depth as $\delta\tau(\bn)=\sigma_T\,n_e\,a\,\delta_e(\bn)$ and the subscript $f$ refers to the standard quadratic estimator filter that minimizes the variance of the estimator subject to the constraint that the estimator is unbiased once applied to the maps as introduced in e.g. Ref.~\citep{Hu:2001kj} and were derived for the quadrupole in Ref.~\citep{Alizadeh:2012vy} for the full sky and in Ref.~\citep{Meyers:2017rtf} for the flat sky, which we omit redriving here. The remote kinetic quadrupole fields measured from the cross of the CMB polarization and galaxies on small scales transform similarly satisfying 
\be
{p^{E}_{\ell m}}' &&=p^{E}_{\ell m} \cos 2 \alpha-p^{B}_{\ell m} \sin 2 \alpha, \\ 
{p^{B}_{\ell m}}'  &&=p^{B}_{\ell m}\cos2\alpha+p^{E}_{\ell m}\sin2\alpha\,.
\ee
The set of observables from cross-correlations of the templates and the reconstructed remote quadrupole are given in Appendix~\ref{sec:app_1} with Eqs.~(A1)-(A10). 

\section{Forecast}\label{sec:forecast}

\subsection{Setup}

In order to assess the detection prospects of cosmic birefringence, we first define an ensemble-information matrix as 
\be
\boldsymbol{\mathcal{F}}_{ab}\!=\!f_{\rm sky}\sum_\ell\frac{(2\ell+1)}{2}{\rm Tr}\!\!\left[\boldsymbol{C}^{-1}\frac{\partial \boldsymbol{C}(\boldsymbol{\ell})}{\partial\boldsymbol{p}_a}\boldsymbol{C}^{-1}\frac{\partial \boldsymbol{C}(\boldsymbol{\ell})}{\partial\boldsymbol{p}_b}\right]\!,\,\,\,\,\,\,
\ee
where $\boldsymbol{C}(\boldsymbol{\ell})$ is the covariance matrix, which we separate into high-$\ell$ and low-$\ell$ contributions, as defined in Appendix~\ref{sec:covariance_matrix}. Note that the low-$\ell$ polarization signal in the CMB is mainly sourced during the epoch of reionization, while the high-$\ell$ spectra are originated at recombination. We calculate the low-$\ell$ spectra for CMB $E$-mode and remote quadrupole field following Refs.~\citep{Deutsch:2017cja,Deutsch:2017ybc}, but when including the contribution from recombination, we take the lensed CMB spectra (including low-$\ell$ $C_\ell^{BB}$) from \textsc{class}~\citep{2011JCAP...07..034B}. We estimate the $1\sigma$-sensitivity to each parameter as~\citep{Tegmark:1996bz}
\be
\sigma_{\boldsymbol{p}_i} = \sqrt{(\boldsymbol{\mathcal{F}}^{-1})_{ii}}\,.
\ee
We assume noise to be diagonal in the covariance matrix. The noises for CMB spectra are given as
\be
N_\ell^{T}&=&(\Delta T)^2 \exp \left[\frac{\ell(\ell+1)\theta_\text{FWHM}^2}{8\ln 2} \right],\label{eq:TT_noise}\\
N_\ell^{E}&=&N_\ell^{B}=2N_\ell^T,
\ee
with the noise amplitudes and angular resolutions given in Table~\ref{tab:noise}. For pSZ tomography, we define 12 bins equally-spaced in comoving distance between ${0.1\le z\le 6}$ and denote by a unit-norm top-hat window function $W(\chi,\chi_i)$ where ${i=\{1,2,\ldots,12\}}$ vary over redshift bins. That is, the remote quadrupole at each bin is given by
\be
p^E_{\ell m, i} = \int\dd \chi\; W(\chi,\chi_i) \;p^E_{\ell m}(\chi),
\ee
and then the angular power spectrum is given by
\be
\Big\langle p^{E*}_{\ell m,i}\;p^E_{\ell'm',j}\Big\rangle = C_\ell^{p^E_i p^E_j}\delta_{\ell\ell'}\delta_{mm'}.
\ee
Using flat-sky approximation, the reconstruction noise for the remote quadrupole at the $i$th bin is given by \cite{Deutsch:2018umo}
\be
&&N_{\ell}^{p^E_i,~p^B_i} \nonumber\\
&&=\left[ \frac{6}{100} \int_{\ell_{\rm min}}^{\ell_{\rm max}} \frac{\ell' d\ell'}{(2\pi)^2} \left( \frac{2\pi}{\sqrt{\tilde{C}_{\ell'}^{EE}\tilde{C}_{\ell'}^{BB}}}\right) \frac{(C_{\ell',i}^{\delta\tau g})^2}{\tilde{C}_{\ell',i}^{gg}}\right]^{-1}\!\!\!\!,\,\,\,\,\,\,
\label{eq:Nq}
\ee
where $\tilde{C}_\ell \equiv C_\ell + N_\ell$, $\{\ell_{\rm min},\ell_{\rm max}\}=\{100,3\times10^4\}$. The binned galaxy-galaxy and optical depth-galaxy power spectra are given by
\be
C^{gg}_{\ell,i} &=& \int \frac{\dd k}{\ell+\frac{1}{2}} W(\chi,\chi_i) b_g(z)^2P_{mm}(k,\chi)\Big|_{\chi\rightarrow\frac{\ell+1/2}{k}},~~~~\\
C^{\delta \tau g}_{\ell ,i} &=& \sigma_T \int \frac{\dd k}{\ell+\frac{1}{2}} (a(\chi) \bar{n}_e(\chi))W(\chi,\chi_i)b_g(z)\nonumber\\
&& \times P_{mm}(k,\chi)\Big|_{\chi\rightarrow\frac{\ell+1/2}{k}},
\ee
where we model the galaxy bias by $b_g(z) = 0.95/D(z)$ where $D(z)$ is the matter growth function obtained using \textsc{camb} \citep{Lewis:1999bs,Howlett:2012mh}, as well as the matter power spectrum $P_{mm}(k,\chi)$. We assumed the electrons trace the total matter so that we simply replace $\delta_e$ with $\delta_m$. We use the same galaxy-galaxy angular power spectrum noise given in \cite{Deutsch:2018umo}
\be
N_{\ell,i}^{gg} = \frac{1}{N_{g,i}}= \left(\int_{\chi_\text{min}}^{\chi_\text{max}}d\chi \;dn(z[\chi]) \right)^{-1},
\ee
where $N_{g,i}$ is the number of galaxies per square radian in the bin and a number density of galaxy which are consistent with VRO is described by
\be
n(z) = n_\text{gal} \frac{1}{2z_0}\left(\frac{z}{z_0}\right)^2\exp(-z/z_0),
\ee
with $n_\text{gal}=40~\text{arcmin}^{-2}$ and $z_0=0.3$.

\begin{table}[t!]
  \centering
  \begin{tabular}{|c|c|c|c|c|c|}
  \hline
    Experiment & Planck & LiteBIRD &  CMB-S4  & CMB-HD & (0.1,0.1) \\
    \hline
    $\Delta T~(\mu{\rm K'})$& $80/\sqrt{2}$ &  $2.16/\sqrt{2}$ &  1 & 0.25 & 0.1 \\ 
    $\theta_{\rm FWHM}$ & $7'$ & $30'$ &  $1.4'$& $0.2'$ & $0.1'$\\
    $\ell$'s & $\ell\!<\!20$& $\ell\!<\!20$ & $\ell\!\geq\!100$ &$\ell\!\geq\!100$ &$\ell\!\geq\!100$  \\
    CMB $C_\ell$'s & $E$,$B$& $E$,$B$ & $T$,$E$,$B$& $T$,$E$,$B$& $T$,$E$,$B$\\
    $f_{\rm sky}$ & 0.5& 0.5 & 0.03 (0.5)& 0.03 (0.5)& 0.03 (0.5)\\
  \hline
  \end{tabular}
  \caption{Survey details used in forecasts. We consider two low-$\ell$ surveys, Planck and LiteBIRD, and three high-$\ell$ surveys, CMB-S4, CMB-HD, and more futuristic hypothetical one which we denote as (0.1,0.1). The low-$\ell$ CMB spectra mainly come from the reionization epoch and the high-$\ell$ spectra pick up the main contributions from the recombination epoch. The $f_{\rm sky}$ values in brackets correspond to the sky fraction assumed for the pSZ tomography.}
\label{tab:noise}
\end{table}

\begin{figure}[t!]
\centering
\includegraphics[width=\linewidth]{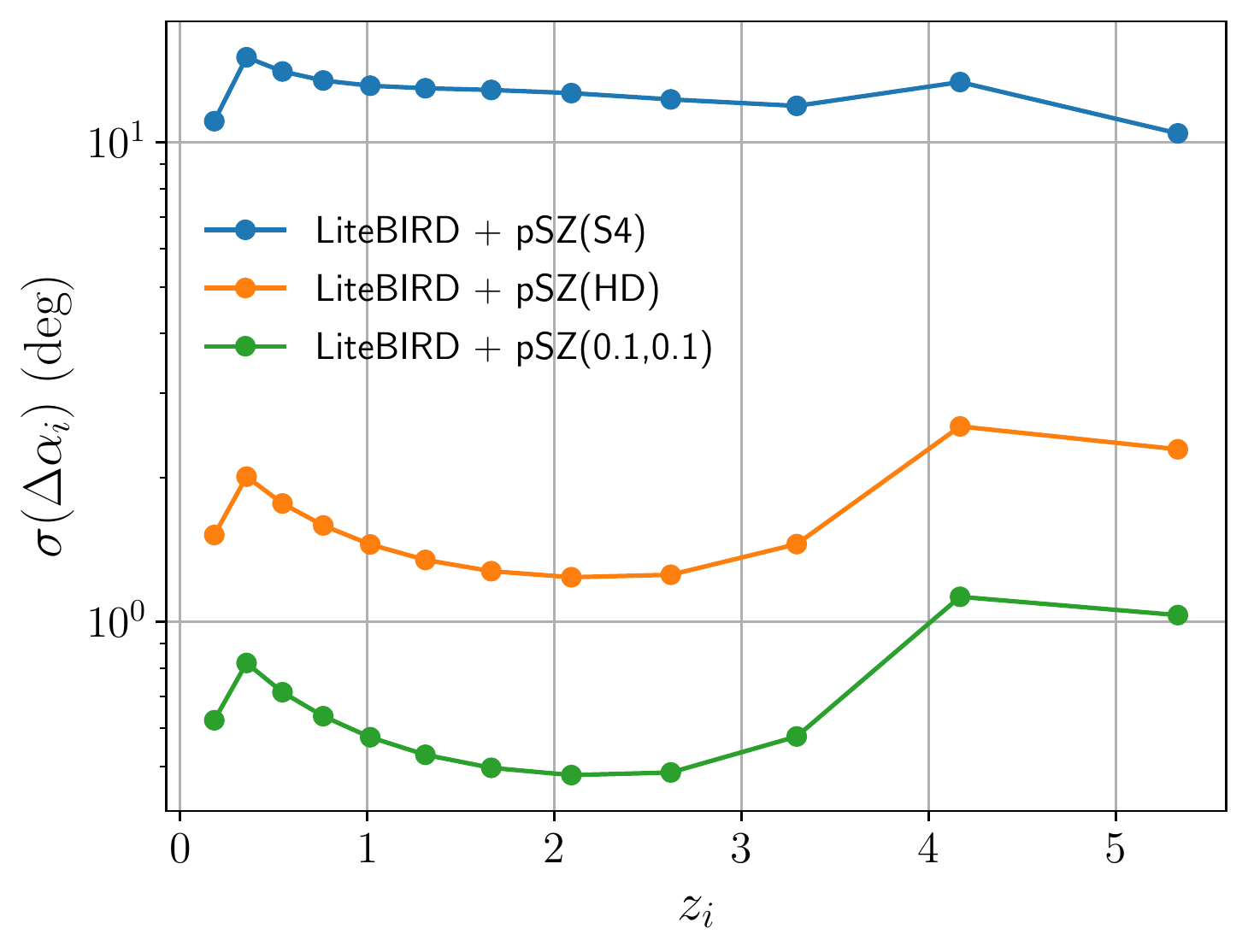}
\caption{The sensitivities on each $\Delta \alpha_i$ which is the change in birefringence angle between two nearby bins. The same angular resolution, $\theta_\text{FWHM}=0.1\text{arcmin}$, is used but with different high-$\ell$ CMB survey. Note that $z_i$ is a median redshift of $i-$th bin. See Sec.~\ref{sec:param1} for the parametrization.}
\label{fig:param1}
\end{figure}

\section{Results}\label{sec:results}

For probing cosmic birefringence, the main advantage of pSZ tomography is that it gives a handle on measuring the rotation angle as a function of redshift. This allows (in addition to providing an additional constraining power) in turn (1) breaking the degeneracy of the signal with the artificial calibration angle $\alpha_{\rm cal}$, (2) allowing birefringence to be probed at times {succeeding} reionization and (3) identifying the redshift range at which the bulk of the birefringence may occur. In what follows, first we demonstrate these advantages by considering four different scenarios for birefringence. Our goal is to highlight the benefits of pSZ tomography both in a generic, model-agnostic way, as well as providing forecasts specific for the axionlike pseudoscalar field introduced above.

\subsection{Intuition}

Our ability to probe cosmic birefringence with pSZ will clearly depend on the fidelity of the CMB and galaxy measurements. In this section, we build intuition for the constraining power of pSZ tomography at different redshift bins and for different experimental choices by considering four simple and demonstrative cosmic birefringence models. 

\subsubsection{Sensitivity at each redshift bin}
\label{sec:param1}
We first define a birefringence angle for each bin as
\be
\alpha_i \equiv \int_{\chi_0}^{\chi_i}\dd\chi\; \frac{\dd\alpha}{\dd\chi},
\ee
where $\chi_i$ is the comoving distance to the bin labeled by $i$, and $\chi_0=0$. Since these angles do not form a set of independent parameters, we use $\Delta \alpha_i$'s  instead as our parameters, which we define as the change in birefringence angle between bin $i$ and $i-1$,
\be
\Delta \alpha_i \equiv \int_{\chi_{i-1}}^{\chi_i}\dd\chi\; \frac{\dd\alpha}{\dd\chi}\,,
\ee
such that each birefringence angle satisfies ${\alpha_i\!=\!\sum_{j=0}^i\Delta \alpha_j}$ and we assume that the birefringence angle from reionization is equivalent to the birefringence angle of the last bin, $\alpha_\text{reio} = \alpha_{12}$. In this way, $\Delta \alpha_i$'s form a set of independent parameters for the information-matrix analysis and enable the pSZ tomography to recover information on the birefringence angle changes as a function of conformal time (or redshift).

Figure \ref{fig:param1} shows the measurement sensitivity to each $\Delta \alpha_i$ from various CMB surveys used for the pSZ tomography, with  reconstruction noise satisfying Eq.~\eqref{eq:Nq}. Despite the relatively weak sensitivity of these surveys when constraining each $\Delta \alpha_i$ individually, the Fig.~\ref{fig:param1}  demonstrates that pSZ tomography is indeed an effective method to study the change in birefringence angle as a function of redshift. In general, the constraining power strongly depends on the number of parameters in the analysis, and one would get better sensitivities if a smaller number of parameters were needed to describe the total birefringence angle, as we explore in what follows.

\begin{figure}[t!]
\centering
\includegraphics[width=\linewidth]{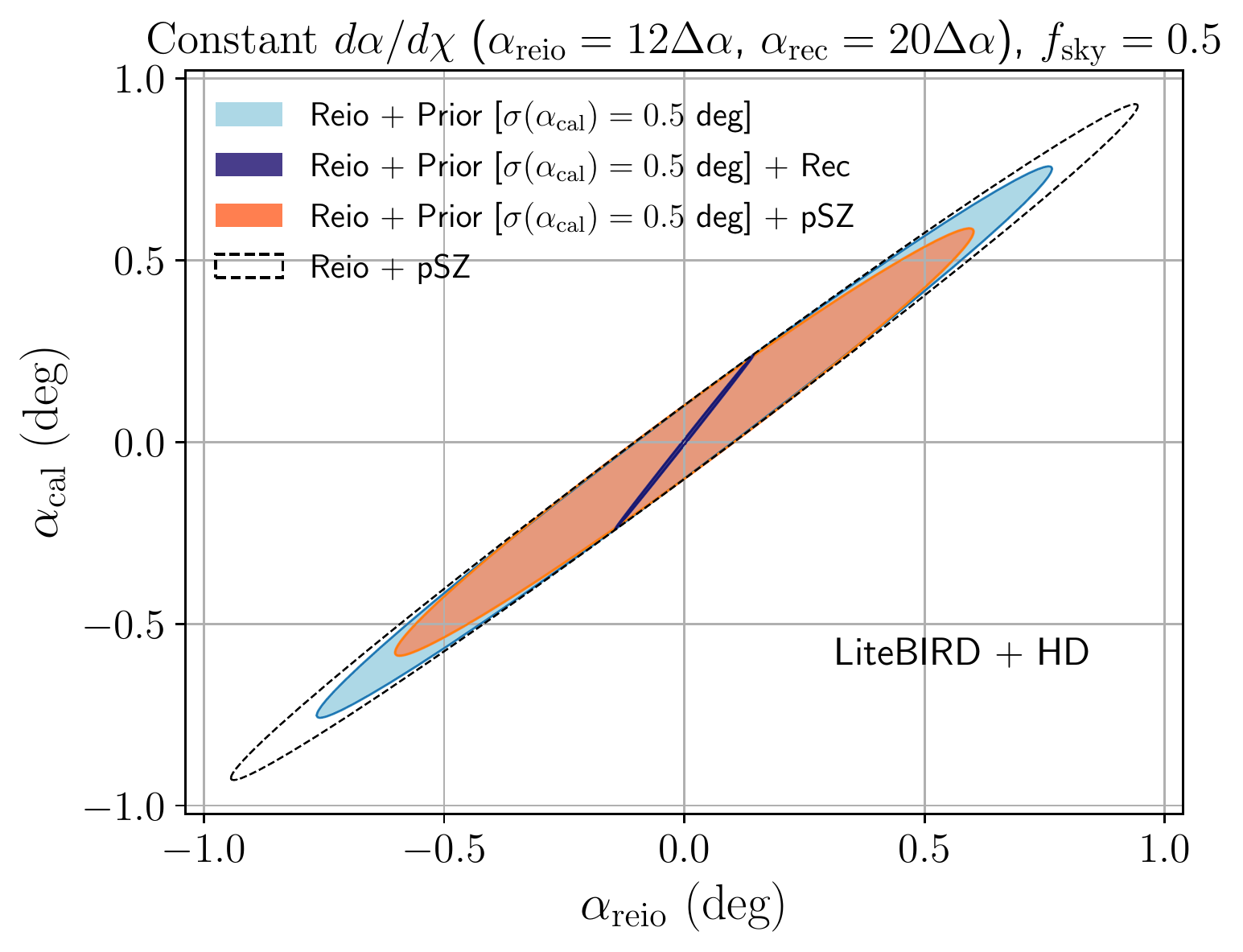}
\includegraphics[width=\linewidth]{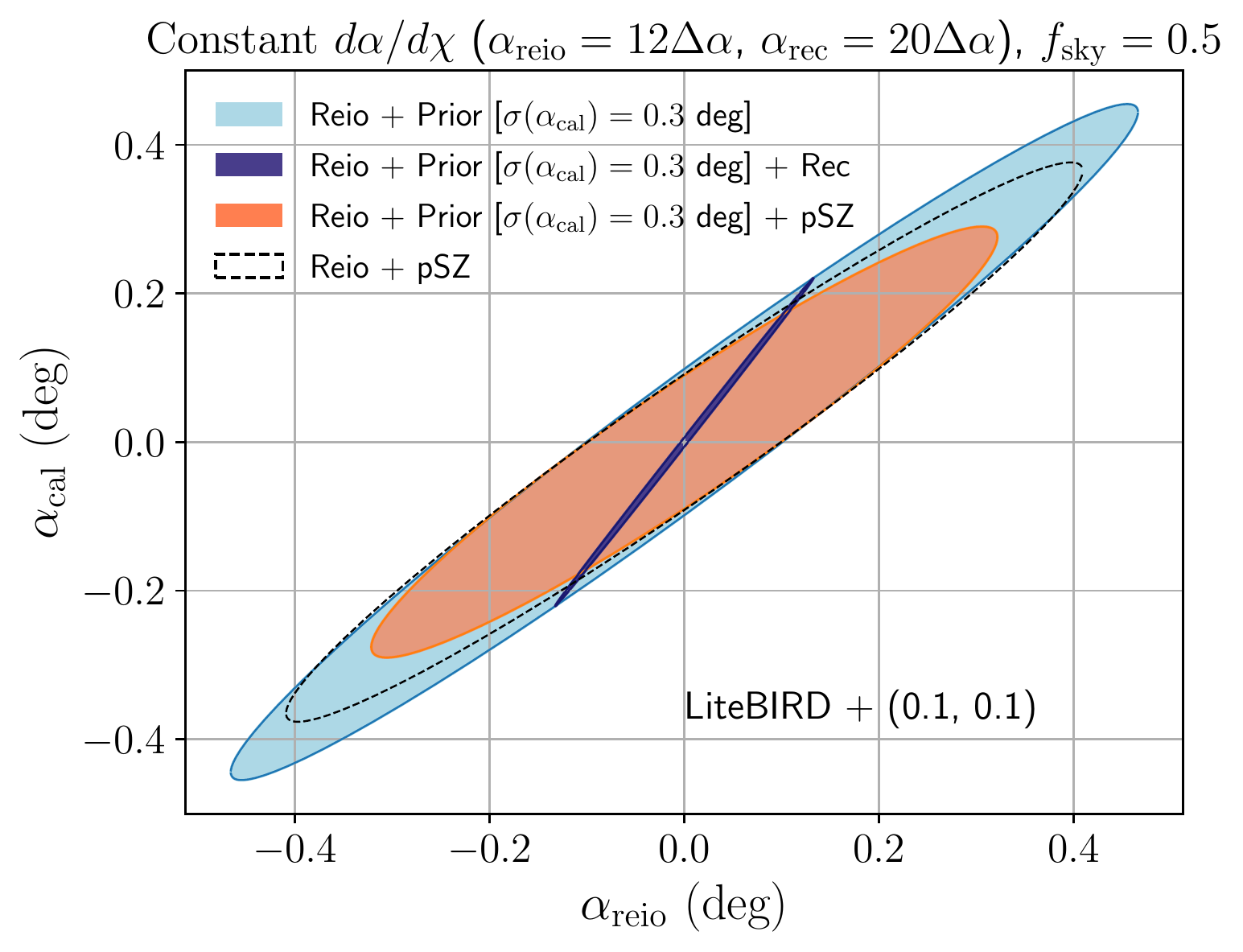}
\caption{Top: LiteBIRD + HD with 0.5 deg prior on the calibration angle $\alpha_{\rm cal}$, bottom: LiteBIRD + (0.1,0.1) with 0.3 deg prior on $\alpha_{\rm cal}$. ``Reio'' (``Rec'') implies the contribution to CMB spectra at low (high) $\ell$'s from reionization (recombination) implies One can benefit from pSZ to break degeneracy between $\alpha_{\rm cal}$ and $\alpha_{\rm reio}$. However, it depends on how tight prior one has on the calibration angle. Note that the total angle is defined as $\alpha_{\rm reio} - \alpha_{\rm cal}.$}
\label{fig:cal-prior}
\end{figure}

\begin{figure}[ht]
\centering
\includegraphics[width=\linewidth]{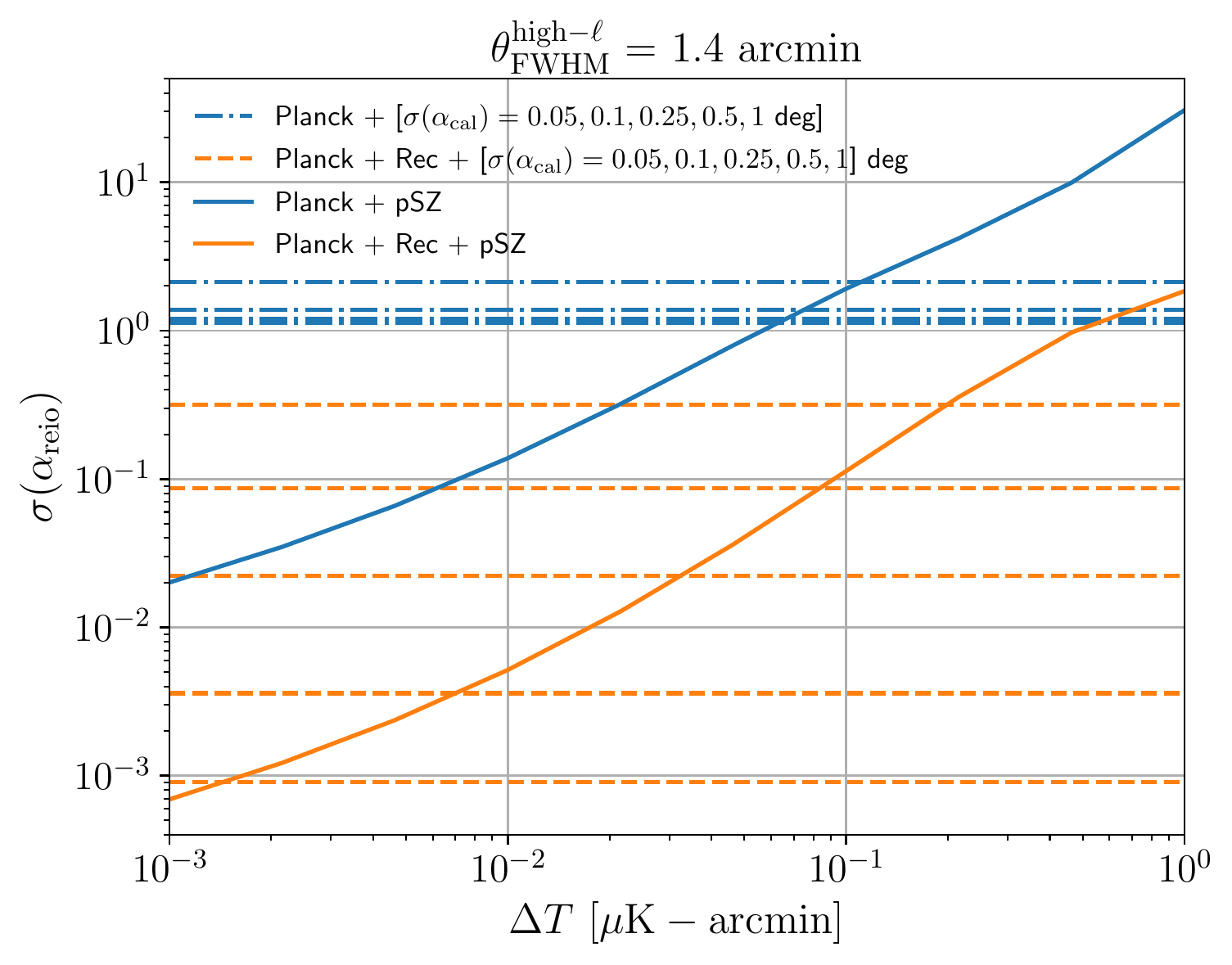}
\includegraphics[width=\linewidth]{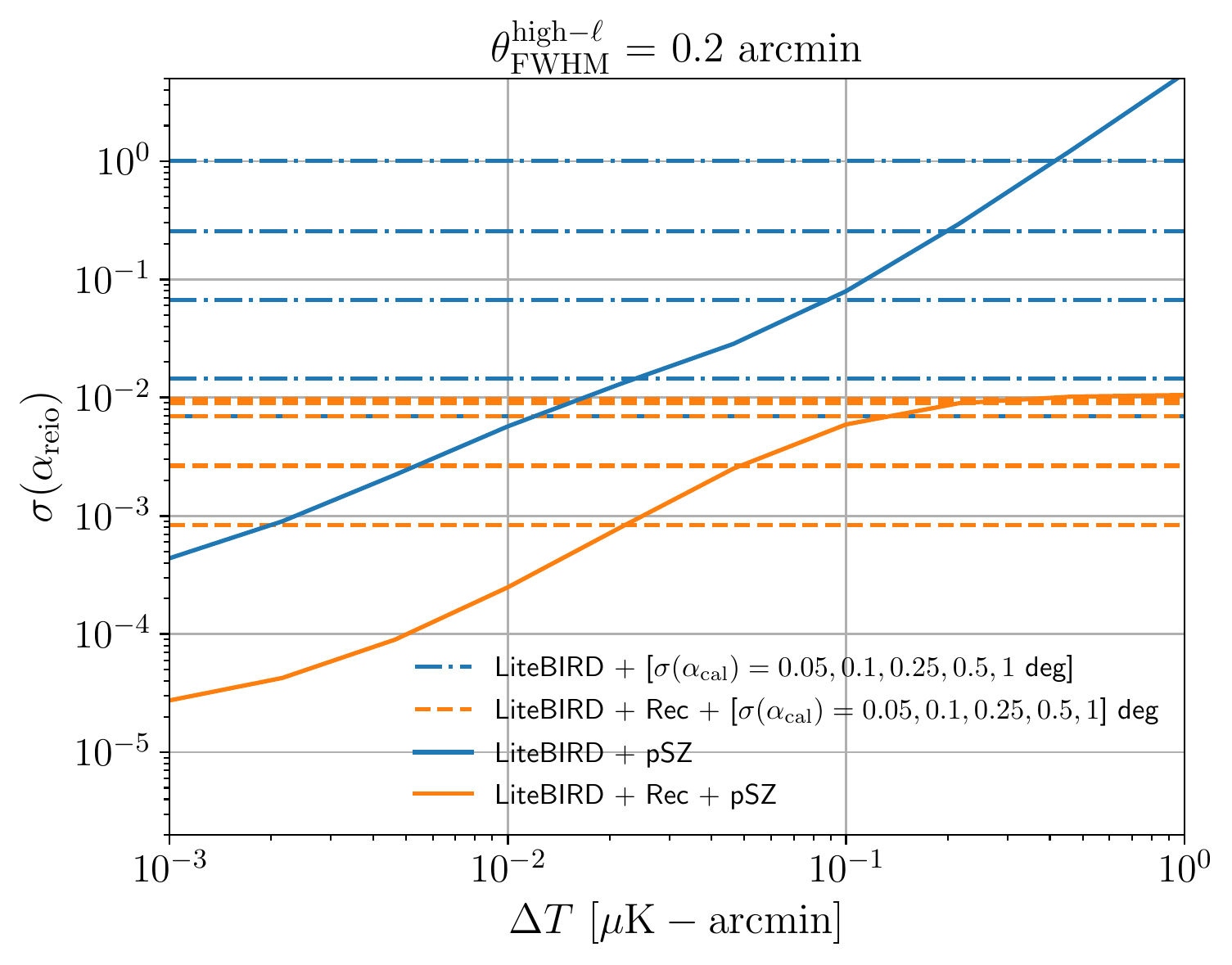}
\caption{A deeper analysis of the prospects of breaking the calibration-angle degeneracy with pSZ-tomography. Solid curves correspond to sensitivities including pSZ tomography and assuming no priors on the calibration angle, as functions of the noise amplitude $\Delta T$ [Eq.~\eqref{eq:TT_noise}]. The dashed curves correspond to including a range of calibration-angle priors without pSZ tomography. Orange (blue) lines include (exclude) recombination signal. The upper (lower) plot assumes Planck (LiteBIRD) measurement of the reionization peak with a CMB-S4 (CMB-HD) like beam for the small-scale CMB polarization used in pSZ tomography. The $x$-axis corresponds to the rms variance of the CMB white noise considered in pSZ tomography, while we fix the recombination CMB noise to forecasted CMB-S4 (CMB-HD) values.}
\label{fig:deeper-dive}
\end{figure}

\subsubsection{Uniform change in the birefringence angle}
\label{sec:param2}

If the physics sourcing birefringence is due to some additional fundamental component of the Universe, it is conceivable that we could model this effect without being totally agnostic to its redshift dependence like we were in the previous section. Here, we explore an additional parametrization which is similar to the one in Sec.~\ref{sec:param1}, but with the simpler assumption that the change in the birefringence angle is constant between two nearby bins,
\be
\Delta \alpha_i \equiv \Delta \alpha = \text{const}\,.
\ee
In this parametrization, the birefringence is defined by a single parameter and the total rotation experienced by photons traveling between reionization and today satisfies $\alpha_\text{reio}= N_\text{bin}\Delta \alpha$, where $N_\text{bin}$ is the number of redshift bins.

For this model, the value of the pSZ tomography is clearer: As demonstrated in Fig.~\ref{fig:cal-prior}, the combination of multiple redshift bins helps breaking the degeneracy between the calibration angle $\alpha_{\rm cal}$ and $\alpha_{\rm reio}$, as the birefringence occurring at different redshifts have different impacts on the reconstructed quadrupole fields and their correlations with the CMB polarization (note that we define $\alpha_{\rm cal}$ to be positively correlated with $\alpha_{\rm reio}$). Furthermore, since the total birefringence experienced by the primary- and reionization-CMB polarization are correlated (they are both factors of $\Delta\alpha$), the combination of these two signals is not sufficient to break the calibration degeneracy completely. The top panel in Fig.~\ref{fig:cal-prior} corresponds to forecasts for a CMB-HD-like experiment, where adding pSZ tomography can be seen to give sensitivities comparable to the combination of the reionization signal with a $0.5^{\circ}$ prior on $\alpha_{\rm cal}$. Combining pSZ-tomography with latter can also be seen to improve the sensitivity. Finally, once a more futuristic survey satisfying $(\Delta T,\theta_{\rm FWHM})=(0.1,0.1)$ is considered (bottom panel), one can benefit even further from pSZ. While combination of reionization and recombination signals lead to better sensitivities compared to pSZ and reionization, for example, the latter is still within a factor 2 of the former, suggesting that pSZ tomography can potentially be used as a validation tool for CMB experiments in the foreseeable future.

Figure \ref{fig:deeper-dive} includes a more detailed analysis of the calibration-angle degeneracy breaking due to pSZ tomography. The solid curves correspond sensitivities to $\alpha_{\rm reio}$ including pSZ tomography, while excluding a calibration angle prior. The horizontal dashed lines correspond to sensitivities from similar observables with a range of calibration-angle priors, while excluding pSZ tomography. The upper (lower) plot corresponds to a Planck-like (LiteBIRD-like) measurement of the reionization peak. In both cases we find pSZ tomography (in combination with reionization and recombination signals) can potentially probe birefringence at a fidelity comparable to having an external measurement of the calibration angle with $\mathcal{O}(0.1)$ degree precision. While it is foreseeable that the CMB spectra from the recombination and reionization epochs can potentially put tight constraints on the total birefringence angle even in the absence of pSZ tomography, here we conjecture that the pSZ tomography can be an alternative validation tool for constraining (or detecting) the cosmic birefringence with the future generation CMB surveys, providing constraints comparable to the former set of observables whose strength will depend on high-precision measurements of the calibration angle.

\begin{figure*}[ht]
\centering
\includegraphics[width=.45\linewidth]{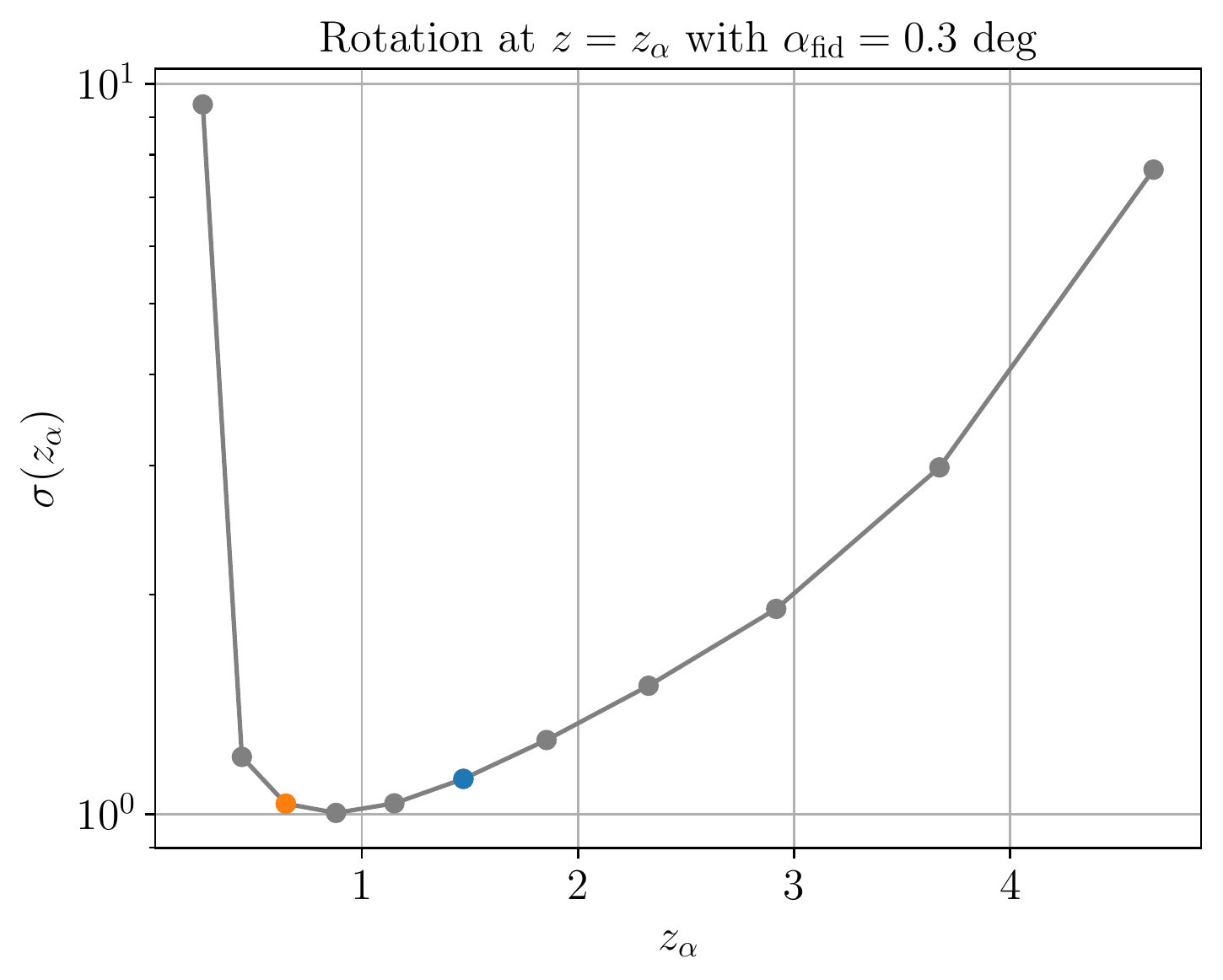}
\includegraphics[width=.45\linewidth]{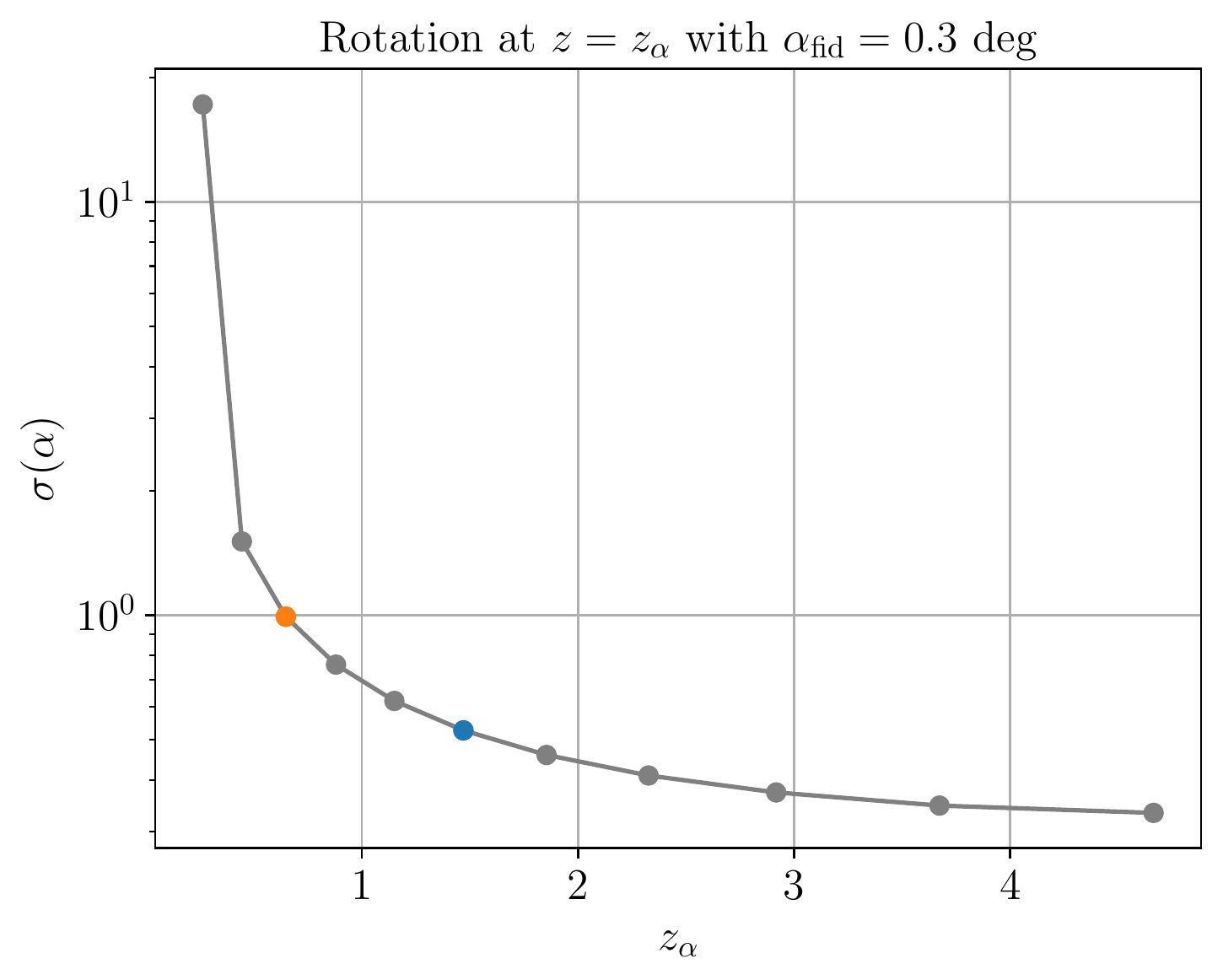}
\includegraphics[width=.45\linewidth]{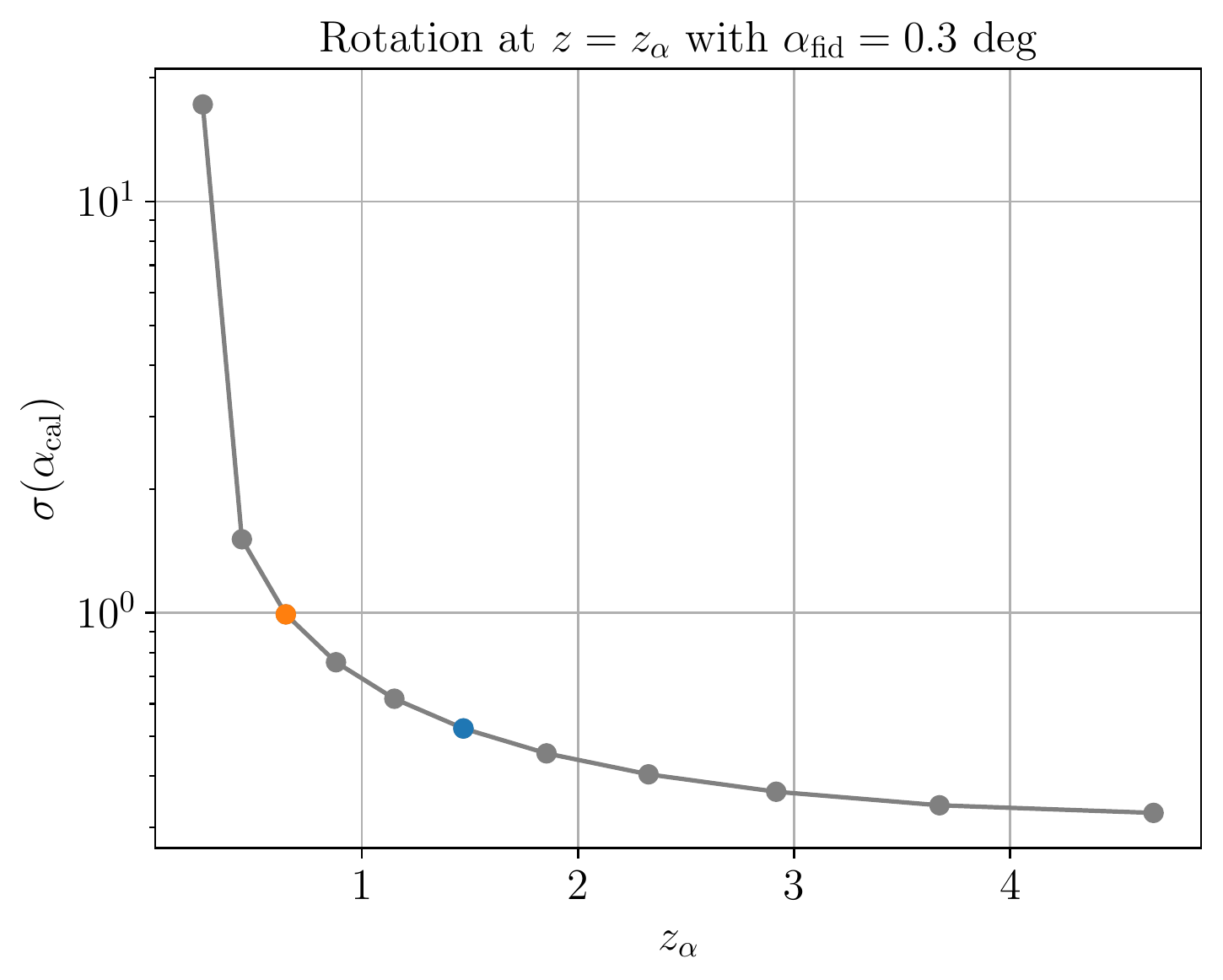}
\includegraphics[width=.45\linewidth]{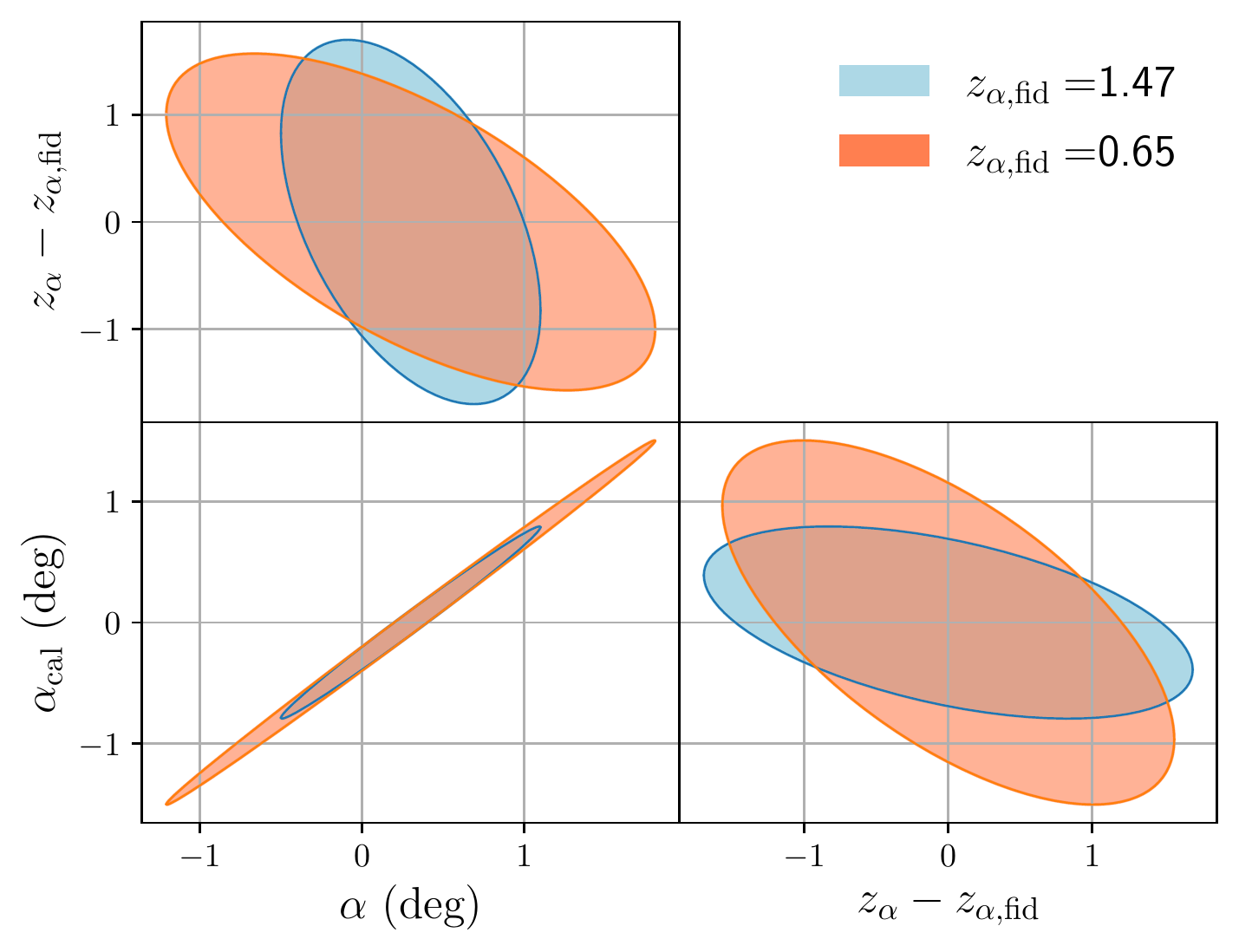}
\caption{Model 1: instant rotation of an angle $\alpha$ at $z=z_\alpha<z_{\rm reio}$ (LiteBIRD + HD), where $z_\alpha$ is the median redshift of each bin. The marginalized sensitivity to each parameter and marginalized contours are shown. Two orange and light blue contours shown in the right bottom panel are corresponding to two bins with median redshifts $z_\alpha=0.65~\text{and}~1.47$, whose marginalized sensitivities in the rest of panels are indicated by the same colors. The sensitivity to $z_\alpha$, which implies how well one can pinpoint the redshift of birefringence, is mainly determined by pSZ signal and that is why, in the top right panel, we see the best sensitivity at bin 4 where the pSZ signal is the biggest. As we go low redshifts, the advantage from pSZ of breaking degeneracy between $\alpha$ and $\alpha_{\rm cal}$ diminishes, hence we see stronger degeneracy between them and weaker sensitivities to them at low redshifts.}
\label{fig:model1}
\end{figure*}

\begin{figure*}[ht]
\centering
\includegraphics[width=.45\linewidth]{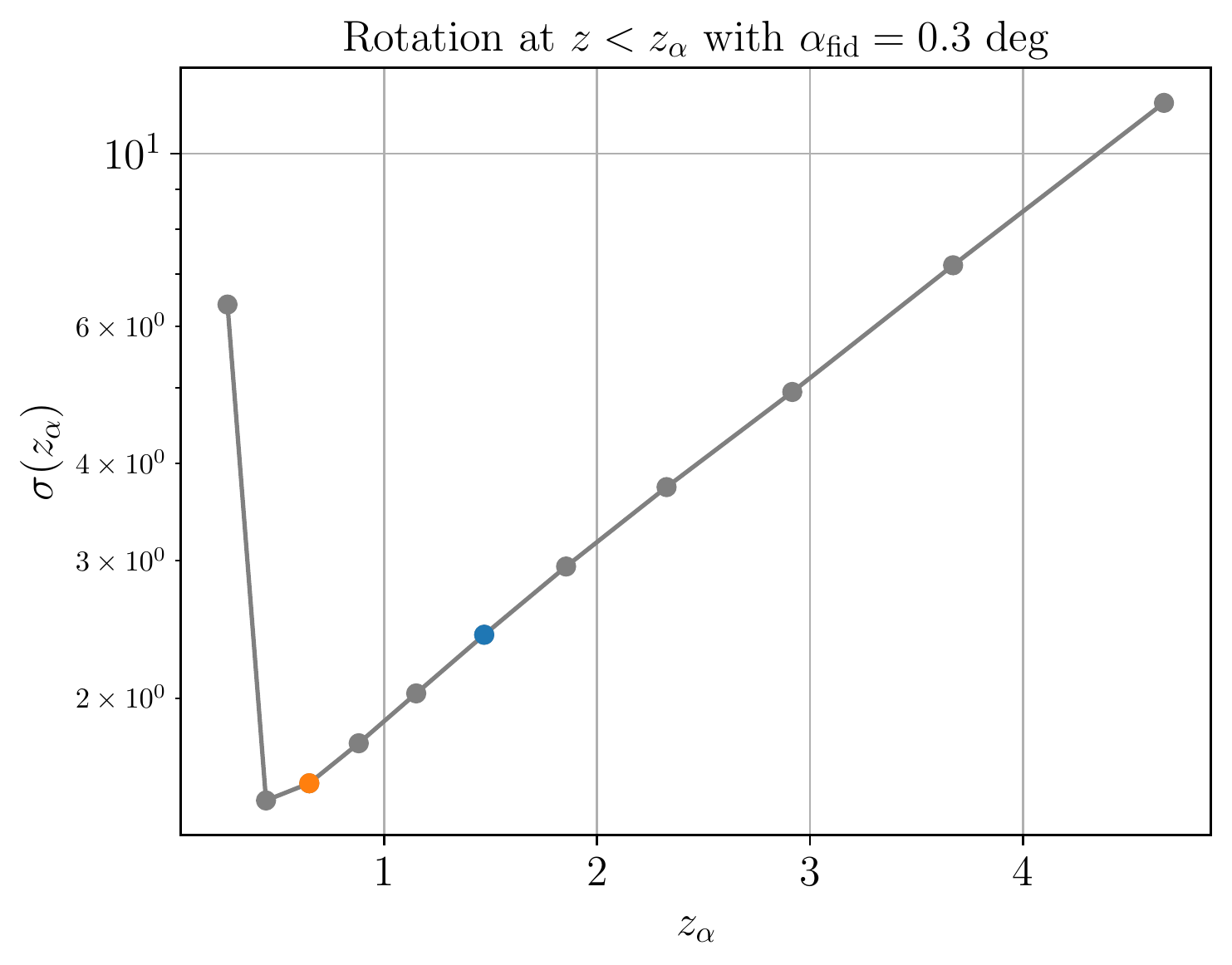}
\includegraphics[width=.45\linewidth]{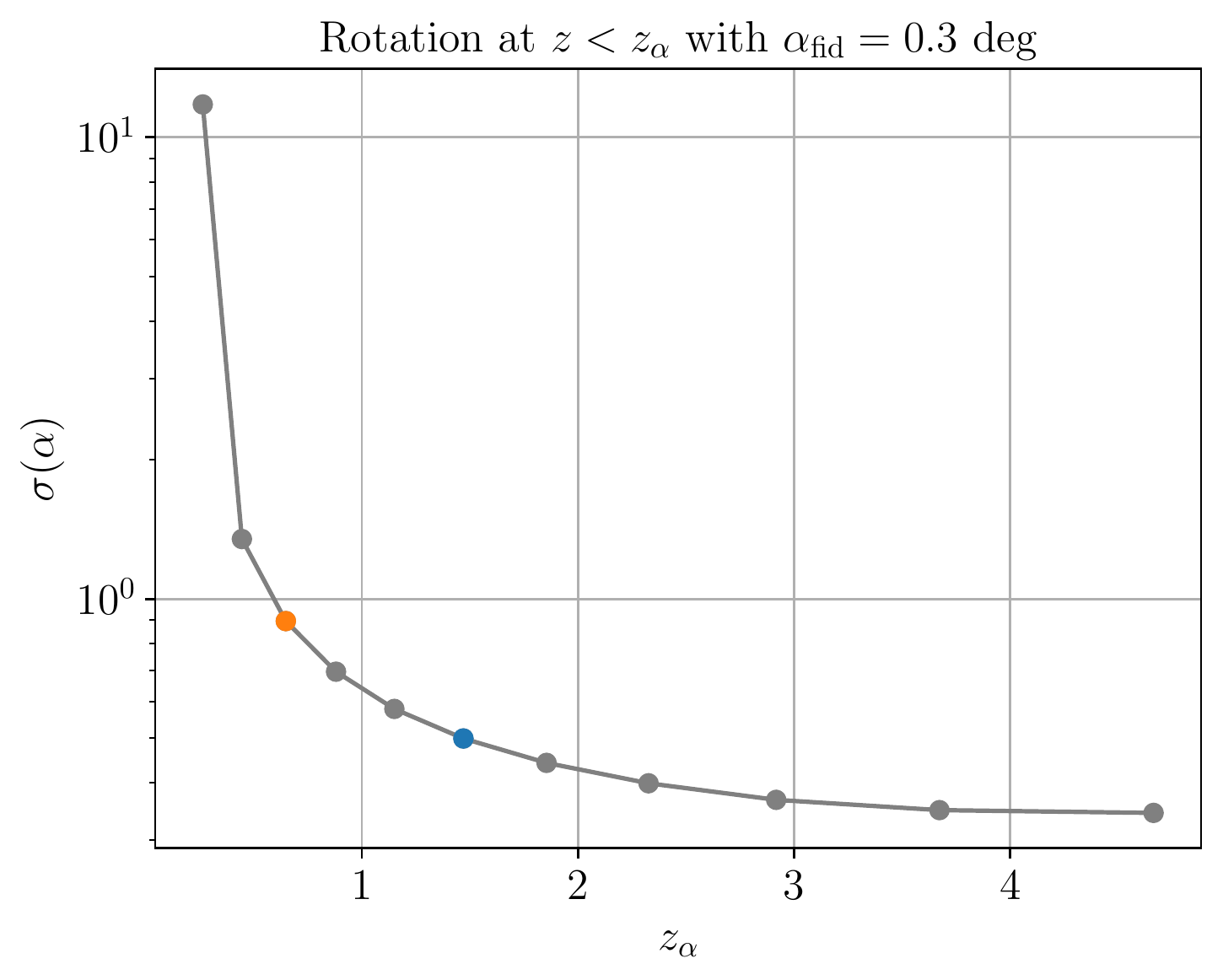}
\includegraphics[width=.45\linewidth]{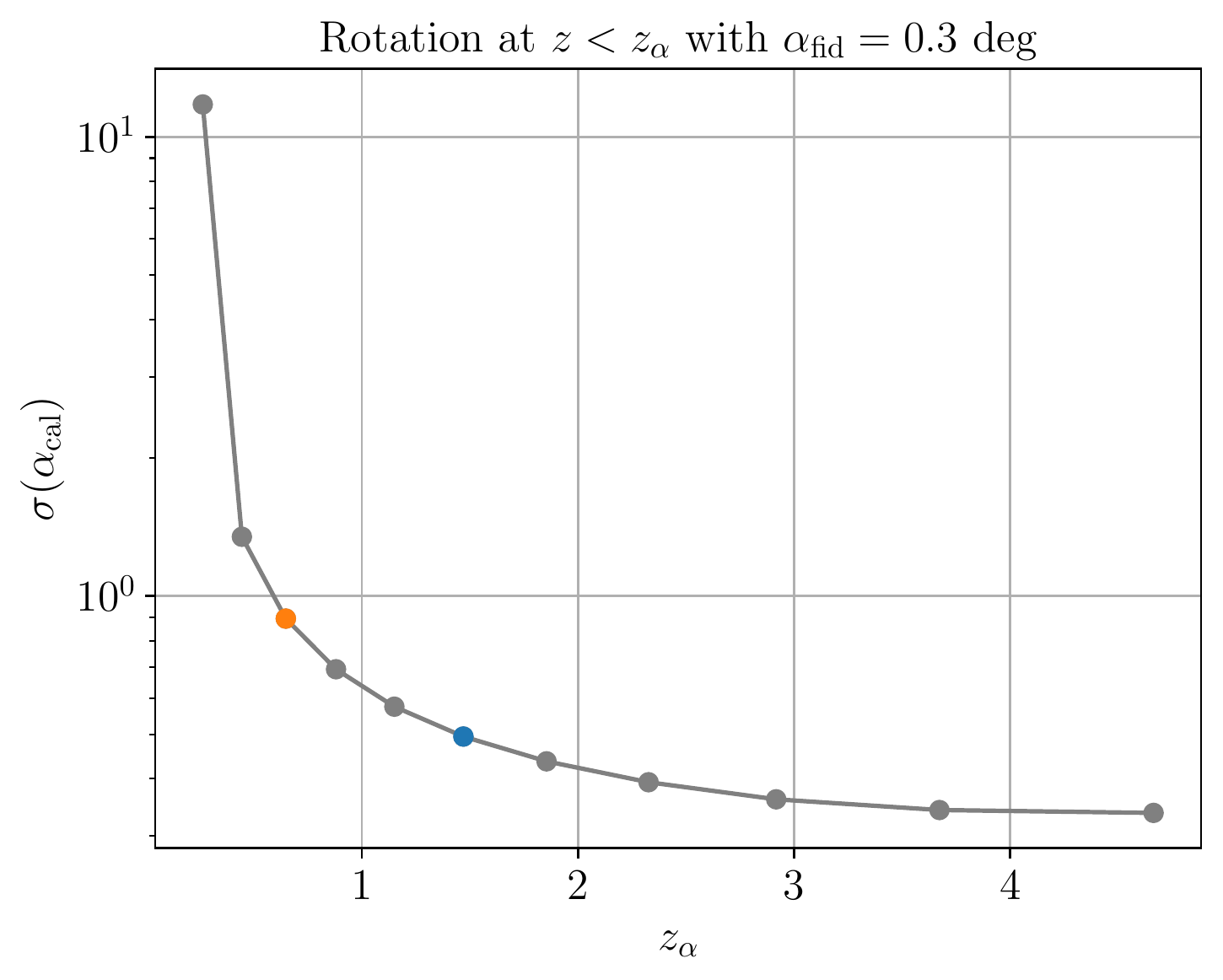}
\includegraphics[width=.45\linewidth]{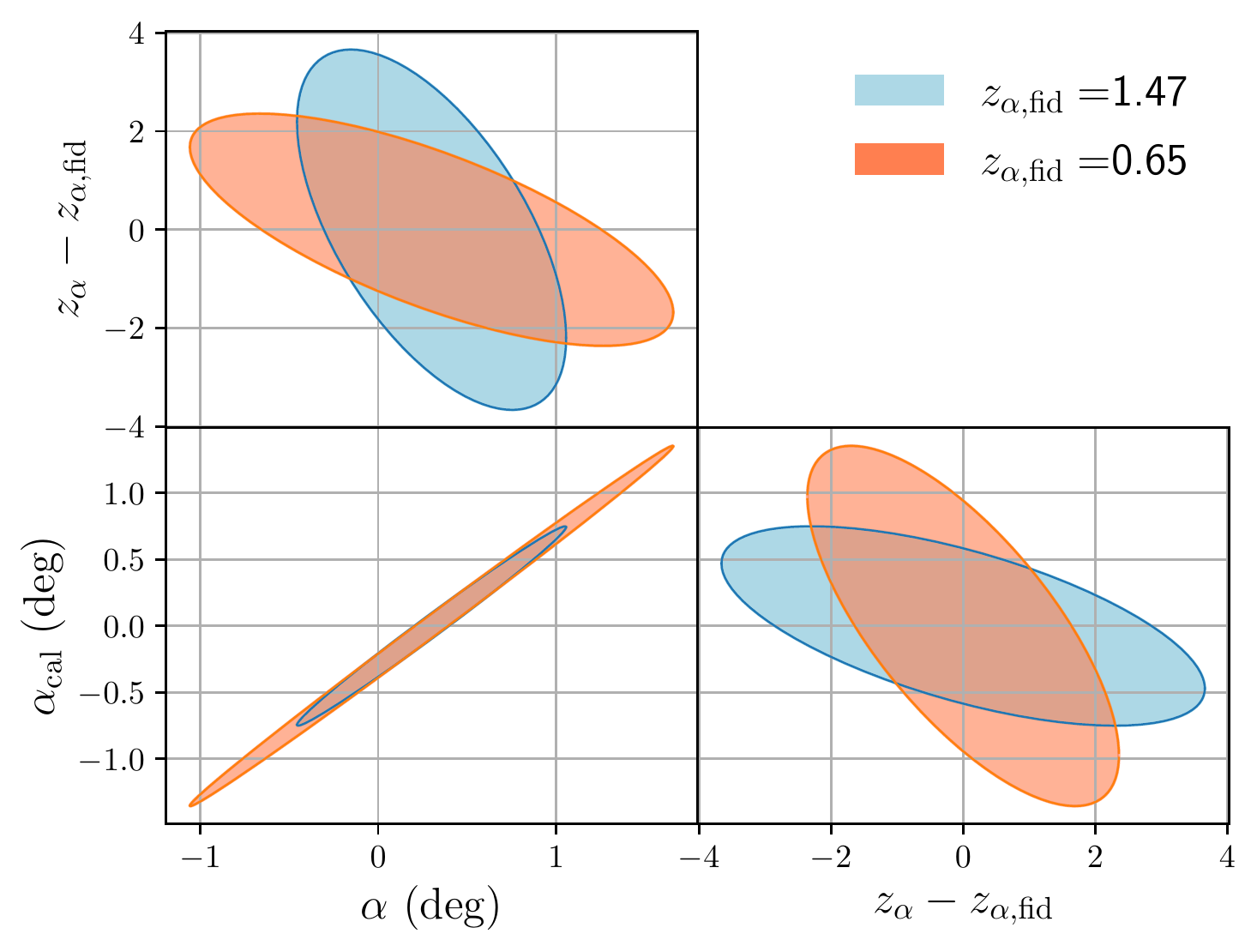}
\caption{The same plots as Fig.~\ref{fig:model1} but with model 2: constant rotation of the total angle $\alpha$ starting at $z = z_\alpha<z_{\rm reio}$ (LiteBIRD + (0.1,0.1)). Note that the total angle $\alpha_\text{fid}=0.3~\text{deg}$ is assumed in each case (each $z_\alpha$) and this results in, unlike model 1, the smaller birefringence angle at each bin as we go high redshifts (when we have more redshift bins going through birefringence).}
\label{fig:model2}
\end{figure*}

\subsubsection{Instant rotation angle $\alpha$ at $z=z_\alpha<z_{\rm reio}$}

Recently, a pseudoscalar ``axionlike" field is considered to explain the observed cosmic birefringence angle at 3$\sigma$ level \cite{Nakatsuka:2022epj}. As this field evolves (or rolls) over its own potential, it rotates the axis of linear polarization. The range of redshifts where the field evolves significantly depends on its mass and potential. Here, and in the following subsection, we first model the field evolution in a simple way to assess the detectability of birefringence from such an evolution of the field. In the next section, we will forecast on the specific scenario of an axionlike pseudoscalar field $\phi$. 

We begin by assuming an instant polarization rotation sourced at some $z=z_\alpha$ with a total ration angle $\alpha$. Such birefringence would impact the polarization produced at larger redshifts $z>z_\alpha$, which corresponds to a scenario where the bulk of the field evolution occurs at a certain redshift $z_\alpha$ in a short timescale. We treat $z_\alpha$ as a free parameter and assume fiducial angle $\alpha_{\rm fid}=0.3^{\circ}$. 

Figure \ref{fig:model1} shows how well pSZ tomography can possibly distinguish the redshift of birefringence occurring at $z_\alpha$ when LiteBIRD and CMB-HD-like CMB experiments are considered. We find that while it may be difficult to detect the redshift of birefringence to high-significance with the next generation of CMB surveys, such as CMB-S4 and Simons Observatory, for such a model, a CMB-HD-like experiment can use pSZ tomography to access this information. Depending on the model, detecting the redshift of birefringence could give a handle on characterizing the source of birefringence. 

\subsubsection{Constant rotation resulting total angle $\alpha$, starting at $z = z_\alpha<z_{\rm reio}$}

Next, we consider a scenario in which the cosmic birefringence occurs throughout $z\le z_\alpha < z_{\rm reio}$ with constant rotation resulting the total angle $\alpha$. This model corresponds to a field which starts evolving at $z=z_\alpha$ and continues evolving until today. We show our results in Fig.~\ref{fig:model2}. 

Having fixed the total angle of rotation to $0.3^\circ$ like before, we find the lower net birefringence at each redshift results in the detectability of the parameters describing this model less promising, even if a LiteBIRD and a futuristic CMB experiment with survey specifications satisfying $\{\Delta T,\theta_{\rm FWHM}\}=(0.1,0.1)$ are considered. 

Nevertheless, we find pSZ tomography can still constrain birefringence  sourced after reionization at the precision of subdegree angle precision as a function of redshift and identify the onset of field evolution to $\mathcal{O}(1)$ precision in redshift in the future. 

\subsection{Probing axionlike dark energy}

As discussed in the previous sections, an axionlike pseudoscalar field $\phi$ with sufficiently small mass $m_\phi$ could constitute all of dark energy and source cosmic birefringence as it evolves through its potential. Reference \citep{Nakatsuka:2022epj} explored the possibility of using the reionization peak and the recombination CMB in combination to probe dark-energy-like axion models. The sensitivity of such a tomographic analysis, however, is limited to birefringence sourced during or before reionization. As shown in Fig.~\ref{fig:phi_evolution}, for lowest axion masses that satisfy the dark-energy-like behavior, the bulk of the field evolution may occur at lowest redshift ranges \textit{after} reionization. Here, we assess the prospects to probing birefringence sourced by a dark-energy-like axion field with mass satisfying $m_\phi<10^{-30}$eV. 

\begin{figure}[ht]
\centering
\includegraphics[width=\linewidth]{LiteBIRD_p2_m_N11.pdf}
\caption{The sensitivities to $m_\phi$ after marginalizing over the total rotation angle $\alpha$ and calibration angle $\alpha_{\rm cal}$. For each $m_\phi$, the rotation angle is normalized to be $0.5^{\circ}$ between recombination and reionization. The LiteBIRD-like survey is assumed for low-$\ell$ spectra and the noise amplitude for high-$\ell$ spectra, which are used to construct the reconstruction noise for pSZ, is varied.}
\label{fig:axion-highz-norm}
\end{figure}

\begin{figure}[ht]
\centering
\includegraphics[width=\linewidth]{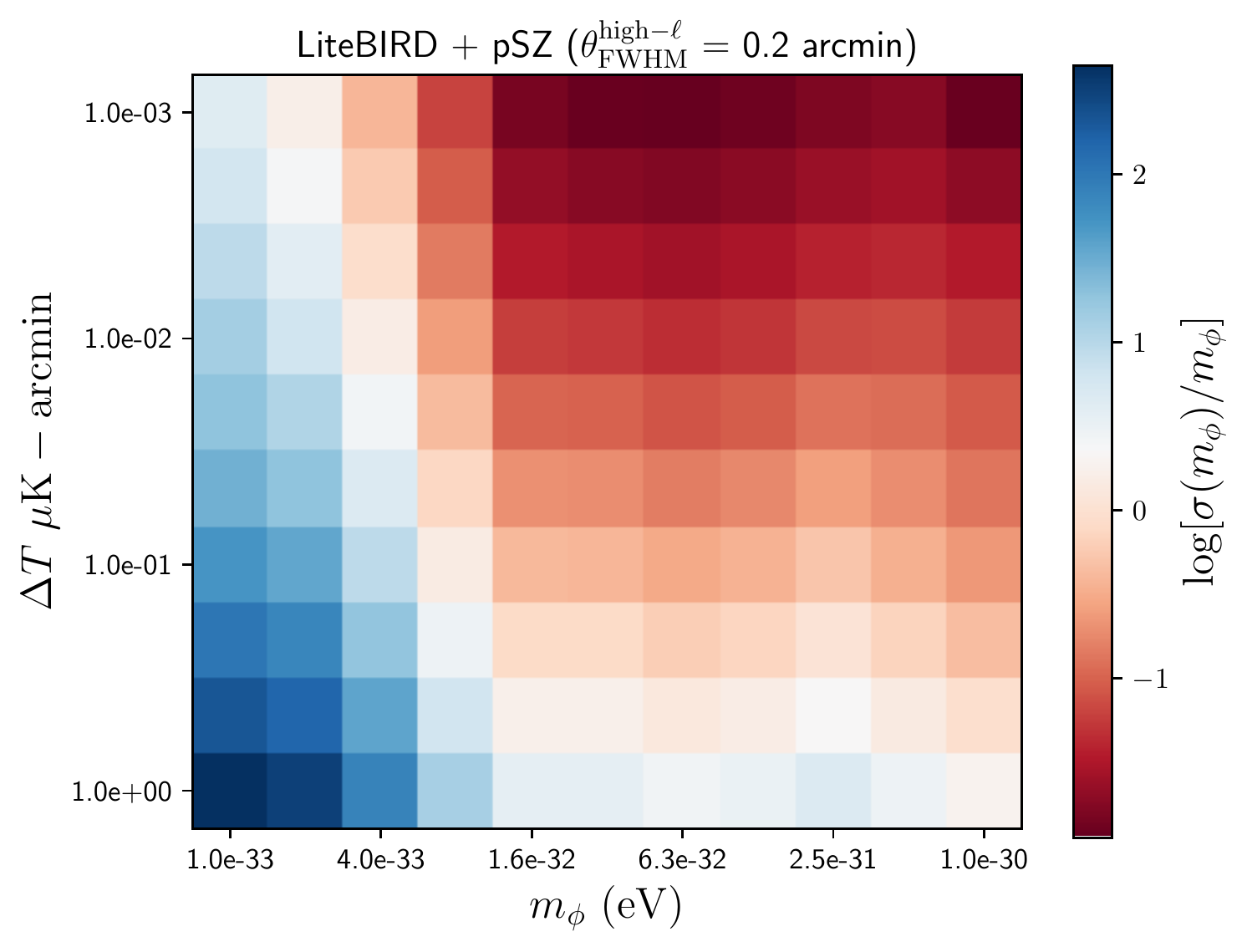}
\caption{The same plot as Fig.~\ref{fig:axion-highz-norm} but the rotation angle is normalized to have the total angle $0.5^{\circ}$.}
\label{fig:axion-total-norm}
\end{figure}

In order to highlight the value of pSZ tomography, we ask the question ``\textit{if we detect evidence for birefringence from comparing the reionization peak to the recombination CMB, could pSZ tomography play a role in ruling out axion models with mass lower than a certain value?}". Since in the absence of pSZ tomography, the effect of birefringence on the combination of reionization and recombination CMB is rather weakly dependent on the axion mass, pSZ tomography can potentially have the role of characterizing the redshift dependence of the rotation, in turn determining the mass of the axion field. Furthermore, for smallest axion masses, observation of some amount of birefringence at redshifts preceding reionization would indicate much larger birefringence at lower redshifts. This would be undetectable from comparing the reionization and recombination CMB signals.   

Figure \ref{fig:axion-highz-norm} demonstrates the prospects to rule out the smallest axion masses with pSZ tomography, if the total rotation sourced between recombination and reionization is $0.5^{\circ}$. As expected, we find that pSZ tomography allows probing models with the smallest axion masses ($m_\phi\lesssim10^{-32}$eV) to high fidelity due to larger total rotation sourced succeeding reionization. On the other hand, in Fig.~\ref{fig:axion-total-norm}, the total rotation sourced between recombination and $z=0$ is set equal to $0.5^{\circ}$. In this latter case, the value of pSZ tomography is less pronounced when constraining the smallest axion masses, as the total rotation is smaller and occurs only very recently, so that pSZ cannot capture the signal. Nevertheless, axion models with larger mass in the range $10^{-32}$--$10^{-30}$eV can be better constrained as the bulk of the rotation is sourced around reionization.

\section{Discussion}\label{sec:discussion}

In this paper, we have explored the prospects to probe cosmic birefringence using pSZ tomography. We have shown the benefits of pSZ tomography both in a generic, model-agnostic way and also by providing forecasts for a specific axionlike pseudoscalar field.

Taking advantage of the ability of measuring the birefringence rotation angle as a function of redshift, we have shown that pSZ tomography can break the degeneracy between the rotation angle $\alpha$ and the artificial calibration angle $\alpha_{\rm cal}$. This implies that pSZ can be considered as a complementary tool for constraining the calibration angle. Here, we simply assumed the constant rate of change of rotation angle $\alpha$ but our calculation could be done with any form of $d\alpha/d\chi$, in general.

We considered two cosmic birefringence models: the one with the instant rotation which occurs after reionization at certain redshift $z=z_\alpha$ and the second one with the constant rotation throughout $z\le z_\alpha$ after reionization. These two models are constructed to mimic some possible pseudoscalar field model, which would cause the possibly detectable cosmic birefringence by pSZ. While the prospects are not very promising, our results show using pSZ tomography could be useful to study the source of birefringence by determining the redshift of rotation.

Finally, we considered the dark-energy-like axion model explored also in Ref.~\citep{Nakatsuka:2022epj} and assessed the scientific value that can be added by the pSZ tomography to a tomographic analysis discussed in Ref.~\citep{Sherwin:2021vgb}, which uses the difference between the recombination and reionization CMB signals to detect and characterize birefringence that is sourced in between. We have found that pSZ tomography may allow probing models with the smallest axion masses $m_\phi\lesssim10^{-32}$eV to high fidelity, if the total rotation between the reionization recombination CMB is $0.5^{\circ}$, for example. 

Note that here we have only considered the ``primary" contributions to the temperature quadrupole that are leading order in perturbation theory, omitting the kinetic pSZ that is recently discussed in Ref.~\citep{Hotinli:2022wbk}. The pSZ signal gets a contribution from the transverse peculiar velocities of electrons which induce a nonzero quadrupole at the electron rest frame that is second order in velocity. While this signal is around an order-of-magnitude smaller than the pSZ signal we consider here, it can still be used to probe fundamental physics as demonstrated in Ref.~\citep{Hotinli:2022wbk}. Nevertheless, we expect the kinetic pSZ to not be a significant source of confusion or bias for the pSZ analysis since it has a frequency dependence different than the CMB black-body and can potentially be removed from maps via ILC-cleaning. 

We have also ignored the optical depth degeneracy due to mismodelling of the small-scale electron-galaxy cross-correlation (as discussed in Ref.~\citep{Smith:2018bpn}, for example). Here, we conjecture that the combination of the kinetic SZ tomography (see e.g.~\citep{Cayuso:2021ljq}), the moving-lens tomography~\citep{Hotinli:2018yyc,Hotinli:2020ntd,Hotinli:2021hih}, as well as the FRB dispersion measurements~\citep{Madhavacheril:2019buy}, will be able to determine the optical depth bias to high precision in the future. 

Overall, we find that the pSZ tomography can potentially increase the measurement quality of cosmic birefringence in the future; increasing the detection signal-to-noise, determining the redshift-dependence of the rotation and providing an alternative way of breaking the calibration-angle degeneracy.

\section{Acknowledgements}

We thank Matthew Johnson for useful conversations and assistance in calculations involving pSZ tomography signal and reconstruction noise. SCH was supported by the Horizon Fellowship from Johns Hopkins University. NL acknowledges the support from New College Oxford/Johns Hopkins Balzan Centre for Cosmological Studies Program.  MK was supported by NSF Grant No.\ 2112699 and the Simons Foundation.

\bibliography{main}

\onecolumngrid
\appendix

\section{Rotated spectra}\label{sec:app_1}

The rotated spectra satisfy
\be
C_\ell^{{{p}^E_i}' {{p}^{E}_j}'}&\simeq& c_i c_j{C_\ell^{{{p}^E_i} {p}_j^{E}}} - c_i s_j\,C_\ell^{{{p}^E_i} {p}_j^{B}}-s_i c_j C_\ell^{{{p}^B_i} {p}_j^{E}}+s_i c_i C_\ell^{{{p}^B_i}{{p}^B_j}},\,\,\,\,\,\,\,\,\, \\
C_\ell^{{{p}^B_i}' {{p}^{B}_j}'}&\simeq& c_i c_j C_\ell^{{{p}^B_i} {p}_j^{B}} + c_i s_j C_\ell^{{{p}^B_i} {p}_j^{E}}+s_i c_j C_\ell^{{{p}^E_i} {p}_j^{B}}+{s_i s_j C_\ell^{{{p}^E_i} {p}_j^{E}}},\,\,\,\,\,\,\,\,\, \\
C_\ell^{{{p}^{B}_i}' {{p}^{E}_j}'}&\simeq&s_i c_j\,C_\ell^{{p}^{E}_i {p}_j^{E}}+c_i c_j C_\ell^{{p}^{B}_i {p}_j^{E}}{-}s_i s_j \,C_\ell^{{p}^{E}_i {p}_j^{B}}-c_i s_j C_\ell^{{p}^{B}_i {p}_j^{B}},\,\,\,\,\,\,\,\,\, \\
C_\ell^{{{p}^{E}_i}' {{p}^{B}_j}'}&\simeq& c_i s_j C_\ell^{{p}^{E}_i {p}_j^{E}}-s_i c_j C_\ell^{{p}^{B}_i {p}_j^{B}}+c_i c_j C_\ell^{{p}^{E}_i {p}_j^{B}}-s_i s_j\,C_\ell^{{p}^{B}_i {p}_j^{E}},\,\,\,\,\,\,\,\,\,\, 
\ee
where we defined $c_i\equiv \cos(2(\alpha_i- \alpha_{\rm cal}))$ and $s_i\equiv \sin(2(\alpha_i-\alpha_{\rm cal}))$, and parametrized the unknown CMB detector calibration with the parameter $\alpha_{\rm cal}$. The cross-correlations with the CMB $E$, $B$ and $T$ modes are
\be
C_\ell^{{E}' {{p}^{E}_i}'}&\simeq&c_{r}c_i{C_\ell^{{E} {p}_i^{E}}}-c_{r}s_iC_\ell^{{E} {p}_i^{B}}-c_i s_r C_\ell^{{B} {p}_i^{E}}+s_r s_i \,C_\ell^{B {p}_j^{B}},\,\,\,\,\,\,\,\,\,\, \\ 
C_\ell^{T {{p}^{E}_i}'}&\simeq&c_i C_\ell^{T {p}_i^{E}}-s_i C_\ell^{T {p}_i^{B}} \,,\\
C_\ell^{{B}'{{p}^{E}_i}'}&\simeq&c_r c_i {C_\ell^{{B} {p}_i^{E}}}-c_r s_i C_\ell^{{B} {p}_i^{B}}{+c_i s_r C_\ell^{{E} {p}_i^{E}}}-s_i s_r \,C_\ell^{E {p}_j^{B}},\,\,\,\,\,\,\,\,\,\, \\
C_\ell^{{E}'{{p}^{B}_i}'}&\simeq&c_r c_i {C_\ell^{{E} {p}_i^{B}}}+c_r s_i  C_\ell^{{E} {p}_i^{E}}-c_i s_r C_\ell^{{B} {p}_i^{B}}-s_r s_i \,C_\ell^{B {p}_j^{E}},\,\,\,\,\,\,\,\,\,\, \\ 
C_\ell^{T {{p}^{B}_i}'}&\simeq&c_i C_\ell^{T {p}_i^{B}}+s_i C_\ell^{T {p}_i^{E}},\,\,\,\,\,\,\,\,\,\, \\
C_\ell^{{B}'{{p}^{B}_i}'}&\simeq&c_r c_i {C_\ell^{{B} {p}_i^{B}}}+c_r s_i C_\ell^{{B} {p}_i^{E}}+c_i s_r C_\ell^{{E} {p}_i^{B}}+s_r s_i \,C_\ell^{E {p}_j^{E}},\,\,\,\,\,\,\,\,\,\,  
\ee
where we define $c_r\equiv \cos(2(\alpha_{\rm reio}\!-\!\alpha_{\rm cal}))$ and $s_r\equiv \sin (2(\alpha_{\rm reio}-\alpha_{\rm cal}))$ with $\alpha_{\rm reio}$ defined as the birefringence experienced by polarization sourced at reionization, and the observed CMB spectra are given by
\be
C_\ell^{E'E'}&& = c_r^2 C_\ell^{EE,\rm reio} + c_{
\rm rec}^2 C_\ell^{EE,\rm rec} +  s_r^2 C_\ell^{BB}+ s_{\rm rec}^2 C_\ell^{BB,\rm rec},~~~~~~\\
C_\ell^{B'B'}&& = c_r^2 C_\ell^{BB,\rm reio} + s_r^2 C_\ell^{EE,\rm reio} + c_{\rm rec}^2 C_\ell^{BB,\rm rec} + s_{\rm rec}^2 C_\ell^{EE,\rm rec},~~~~~~ \\
C_\ell^{T'E'}&& = c_r C_\ell^{TE,\rm reio}+ c_{\rm rec} C_\ell^{TE, \rm rec}, \\
C_\ell^{T'B'} &&=  s_{r}C_\ell^{TE}+s_{\rm rec}C_\ell^{TE,\rm rec} ,\\
C_\ell^{E'B'} &&=  s_r c_r (C_\ell^{EE,\rm reio}-C_\ell^{BB,\rm reio})+ s_{\rm rec}c_{\rm rec} (C_\ell^{EE,\rm rec}-C_\ell^{BB,\rm rec}).
\ee

\section{Covariance Matrix}\label{sec:covariance_matrix}

The low-$\ell$ contribution to the covariance matrix satisfy, 
\begin{widetext}
\begin{equation}
\boldsymbol{C}^{{\rm low}-\ell}=\begin{pmatrix}
C_\ell^{E'E'} &
C_\ell^{E'B'} & 
C_\ell^{{{p}^E_1}'{E'}} &
C_\ell^{{{p}^E_2}'{E'}} & 
\ldots &
C_\ell^{{{p}^E_N}'{E'}} &
C_\ell^{{{p}^B_1}'{E'}} &
C_\ell^{{{p}^B_2}'{E'}} &
\ldots &
C_\ell^{{{p}^B_N}'{E'}}\\
C_\ell^{B'E'} &
C_\ell^{B'B'} & 
C_\ell^{{{p}^E_1}'{B'}} & 
C_\ell^{{{p}^E_2}'{B'}} & 
\ldots &
C_\ell^{{{p}^E_N}'{B'}} &
C_\ell^{{{p}^B_1}'{B'}} &
C_\ell^{{{p}^B_2}'{B'}} &
\ldots &
C_\ell^{{{p}^B_N}'{B'}}\\
C_\ell^{{{p}^E_1}'{E}'} &
C_\ell^{{{p}^E_1}'{B}'} & 
C_\ell^{{{p}^E_1}'{{p}^E_1}'} & C_\ell^{{{p}^E_2}'{{p}^E_1}'} & 
\ldots &
C_\ell^{{{p}^E_N}'{{p}^E_1}'} &
C_\ell^{{{p}^B_1}'{{p}^E_1}'} &
C_\ell^{{{p}^B_2}'{{p}^E_1}'} &
\ldots &
C_\ell^{{{p}^B_N}'{{p}^E_1}'}\\
C_\ell^{{{p}^E_2}'{E}'} &
C_\ell^{{{p}^E_2}'{B}'} & 
C_\ell^{{{p}^E_1}'{{p}^E_2}'} & C_\ell^{{{p}^E_2}'{{p}^E_2}'} & 
\ldots &
C_\ell^{{{p}^E_N}'{{p}^E_2}'} &
C_\ell^{{{p}^B_1}'{{p}^E_2}'} &
C_\ell^{{{p}^B_2}'{{p}^E_2}'} &
\ldots &
C_\ell^{{{p}^B_N}'{{p}^E_2}'}\\
\vdots & \vdots & \ddots & 
\vdots & \vdots &\vdots 
& \ddots & \vdots \\
C_\ell^{{{p}^E_N}'{E}'} &
C_\ell^{{{p}^E_N}'{B}'} & 
C_\ell^{{{p}^E_1}'{{p}^E_N}'} & C_\ell^{{{p}^E_2}'{{p}^E_N}'} & 
\ldots &
C_\ell^{{{p}^E_N}'{{p}^E_N}'} &
C_\ell^{{{p}^B_1}'{{p}^E_N}'} &
C_\ell^{{{p}^B_2}'{{p}^E_N}'} &
\ldots &
C_\ell^{{{p}^B_N}'{{p}^E_N}'}\\
C_\ell^{{{p}^B_1}'{E}'} &
C_\ell^{{{p}^B_1}'{B}'} & 
C_\ell^{{{p}^E_1}'{{p}^B_1}'} & C_\ell^{{{p}^E_2}'{{p}^B_1}'} & 
\ldots &
C_\ell^{{{p}^E_N}'{{p}^B_1}'} &
C_\ell^{{{p}^B_1}'{{p}^B_1}'} &
C_\ell^{{{p}^B_2}'{{p}^B_1}'} &
\ldots &
C_\ell^{{{p}^B_N}'{{p}^B_1}'}\\
C_\ell^{{{p}^B_2}'{E}'} &
C_\ell^{{{p}^B_2}'{B}'} & 
C_\ell^{{{p}^E_1}'{{p}^B_2}'} & C_\ell^{{{p}^E_2}'{{p}^B_2}'} & 
\ldots &
C_\ell^{{{p}^E_N}'{{p}^B_2}'} &
C_\ell^{{{p}^B_1}'{{p}^B_2}'} &
C_\ell^{{{p}^B_2}'{{p}^B_2}'} &
\ldots &
C_\ell^{{{p}^B_N}'{{p}^B_2}'}\\
\vdots &\vdots &\ddots & 
\vdots &\vdots & \vdots 
& \ddots & \vdots \\
C_\ell^{{{p}^B_N}'{E}'} &
C_\ell^{{{p}^B_N}'{B}'} & 
C_\ell^{{{p}^E_1}'{{p}^B_N}'} & C_\ell^{{{p}^E_2}'{{p}^B_N}'} & 
\ldots &
C_\ell^{{{p}^E_N}'{{p}^B_N}'} &
C_\ell^{{{p}^B_1}'{{p}^B_N}'} &
C_\ell^{{{p}^B_2}'{{p}^B_N}'} &
\ldots &
C_\ell^{{{p}^B_N}'{{p}^B_N}'}
\end{pmatrix},
\label{eq:cov}
\end{equation}
\end{widetext}
where the high-$\ell$ contribution is
\be
\boldsymbol{C}^{{\rm high}-\ell}= \begin{pmatrix}
C_\ell^{TT} & C_\ell^{TE'} & C_\ell^{TB'}\\
C_\ell^{E'T} & C_\ell^{E'E'} & C_\ell^{E'B'} \\
C_\ell^{B'T} & C_\ell^{B'E'} & C_\ell^{B'B'} 
\end{pmatrix}.
\ee

\begin{widetext}
\section{The remote quadrupole moment seen by electrons}\label{sec:app_c}

Here we provide the equations used in this work following \citep{Deutsch:2017cja,Deutsch:2017ybc,Deutsch:2018umo}. As shown in Eq.~\eqref{eq:Theta}, there are three contributions (SW, ISW, Doppler effect) to the local CMB temperature perturbations, which are given by
\be
\Theta_{\rm SW} (\chi_e, \bn_e,\bn) &=& \left( 2D_\Psi(\chi_{\rm dec}) - \frac{3}{2}\right) \Psi_i(\br_{\rm dec}),\\
\Theta_{\rm ISW}(\chi_e, \bn_e,\bn) &=& 2 \int_{a_{\rm dec}}^{a_e} \frac{\dd D_\psi}{\dd a} \Psi_i(\br(a))da,\\
\Theta_{\rm Doppler}(\chi_e, \bn_e,\bn) &=& \bn \cdot [D_v(\chi_{\rm dec})\nabla \Psi_i(\br_{\rm dec})-D_v(\chi_e)\nabla \Psi_i(\br_e)],
\ee
where $\br(a) \equiv \chi_e \bn_e + \Delta \chi(a)\bn$, $\Delta \chi(a) = - \int_{a_e}^a da' [H(a') a'^2]^{-1}]$ and $D_\psi(\chi)$ is the growth function of the potential defined as
\be
\Psi(\br,\chi) = D_\psi(\chi)\Psi_i(\br).
\ee
This growth function and the velocity growth function $D_v$ defined by $\bv (\br,\chi) = -D_v(\chi) \nabla \Psi_i(\br)$ are given by
\be
D_\Psi(a) &=& \frac{ 16 \sqrt{1+y} + 9y^3 + 2y^2 - 8y - 16}{10y^3} \left[ \frac{5\Omega_m }{2} \frac{H(a)}{aH_0} \int_0^a \frac{da'}{(a'H(a')/H_0)^3}\right],\\
D_v(a) &=& \frac{2a^2H(a)}{H_0^2\Omega_m} \frac{y}{4+3y} \left[D_\Psi(a) + \frac{\dd D_\Psi(a)}{\dd \ln a}\right],
\ee
where $y\equiv a/a_{\rm eq}$. By writing the primordial potential $\Psi_i$ in Fourier space, the local CMB quadrupole can be rewritten as Eq.~\eqref{eq:Theta_2m},
\be
\Theta_{2m}(\chi_e\bn_e) &=& \int \frac{\dd^3k}{(2\pi)^3} \Psi_i(\bm{k}) T(k) [ \mathcal{G}_{\rm SW}(k,\chi_e)+\mathcal{G}_{\rm ISW}(k,\chi_e) + \mathcal{G}_{\rm Doppler}(k,\chi_e)] Y_{2m}^*(\khat)\;e^{i\chi_e\bsk \cdot \bn_e},
\ee
where the kernels are given by
\be
\mathcal{G}_{\rm SW}(k,\chi_e) &=& -4\pi \left( 2D_\Psi(\chi_{\rm dec})-\frac{3}{2}\right) j_2(k\Delta \chi_{\rm dec}),\\
\mathcal{G}_{\rm ISW}(k,\chi_e) &=& - 8\pi \int_{a_{\rm dec}}^{a_e} \dd a\; \frac{\dd D_\Psi}{\dd a}\; j_2(k\Delta \chi(a)),\\
\mathcal{G}_{\rm Doppler}(k,\chi_e) &=& \frac{4\pi}{5} kD_v(\chi_{\rm dec})[3j_3(k\Delta \chi_{\rm dec}) - 2j_1(k\Delta \chi_{\rm dec})],
\ee
and the transfer funcion $T(k)$ is given by the Bardeen-Bond-Kaiser-Szalay fitting function \citep{Bardeen:1985tr},
\be
T(k) = \frac{\ln[1+0.171x]}{0.171x} [ 1+0.284x + (1.18x)^2 + (0.339x)^3 + (0.49x)^4]^{-0.25},
\ee
where $x=k/k_{\rm eq}$ and $k_{\rm eq} = a_{\rm eq}H(a_{\rm eq})$. Then, using Eqs.~\eqref{eq:remote_quad} and \eqref{eq:Theta_2m}, the multipole coefficients can be obtained by
\be
p^E_{\ell m}(\chi)= \int \dd^2 \bn_e \; {}_\pm p(\chi_e \bn_e) {}_{\pm2} Y^*_{\ell m}(\bn_e)
= \int \frac{\dd^3k}{(2\pi)^3} \Delta_\ell^p(\chi, k) \Psi_i(\bsk) Y_{\ell m}^*(\khat),
\ee
where the remote quadrupole transfer function $\Delta_\ell^p(\chi_e,k)$ is
\be
\Delta_\ell^p(\chi_e,k) &=& -5i^\ell \sqrt{\frac{3}{8}} \sqrt{\frac{(\ell+2)!}{(\ell-2)!}} \frac{j_\ell(k\chi_e)}{(k\chi_e)^2} \;T(k)
 \;[ \mathcal{G}_{\rm SW}(k,\chi_e)+\mathcal{G}_{\rm ISW}(k,\chi_e) + \mathcal{G}_{\rm Doppler}(k,\chi_e)].
\ee

\end{widetext}

\end{document}